\documentclass[iop]{emulateapj} % use larger type; default would be 10pt
\usepackage{multirow}
\usepackage{natbib}

\newcommand{\um}{$\mu$m}
\newcommand{\msun}{\mbox{M$_\odot$}}% Msun
\newcommand{\lsun}{\mbox{L$_\odot$}}% Lsun
\newcommand{\tbol}{\mbox{$T_{bol}$}} % bolometric temperature
\newcommand{\lsmm}{\mbox{$L_{smm}$}} % bolometric temperature
\newcommand{\lbolsmm}{\mbox{$L_{bol}/L_{smm}$}} % bol-to-smm luminosity
\newcommand{\lsmmbol}{\mbox{$L_{smm}/L_{bol}$}} % smm-to-bol luminosity
\newcommand{\mjybeam}{mJy~beam$^{-1}$}

\usepackage{epstopdf}
\slugcomment{\footnotesize Resubmitted to AJ on \today}

\begin{document}
\LongTables
%\Maketitle

\title {A Catalog of Low-Mass Star-Forming Cores Observed with SHARC-II at 350 microns}

\author{
Akshaya Suresh\altaffilmark{1}, 
Michael M.~Dunham\altaffilmark{2,3}, 
H\'ector G.~Arce\altaffilmark{1}, 
Neal J. Evans II\altaffilmark{4},
Tyler L.~Bourke\altaffilmark{2,5},
Manuel Merello\altaffilmark{6}, \& 
Jingwen Wu\altaffilmark{7}
}

\altaffiltext{1}{Department of Astronomy, Yale University, P.O. Box 208101, New Haven, CT 06520, USA}

\altaffiltext{2}{Harvard-Smithsonian Center for Astrophysics, 60 Garden Street, MS 78, Cambridge, MA 02138, USA}

\altaffiltext{3}{E-mail address: mdunham@cfa.harvard.edu}

\altaffiltext{4}{Department of Astronomy, The University of Texas at Austin, 2515 Speedway, Stop C1400, Austin, TX 78712-1205, USA}

\altaffiltext{5}{SKA Organization, Jodrell Bank Observatory, Lower Withington, Macclesfield, Cheshire SK11 9DL, UK}

\altaffiltext{6}{Istituto di Astrofisica e Planetologia Spaziali-INAF, Via Fosso del Cavaliere 100, I-00133 Roma, Italy}

\altaffiltext{7}{National Astronomical Observatories, Chinese Academy of Sciences, 20A Datun Road, Chaoyang District, Beijing, China, 100012}

\begin{abstract}
We present a catalog of low-mass dense cores observed with 
the SHARC-II instrument at 350 $\mu$m. Our observations have 
an effective angular resolution of 10$''$, approximately 2.5 times higher 
than observations at the same wavelength obtained with the 
{\it Herschel Space Observatory}, albeit with lower sensitivity, especially 
to extended emission. The catalog includes 81 maps covering 
a total of 164 detected sources. For each detected source, we tabulate basic 
source properties including position, peak intensity, flux density in fixed 
apertures, and radius.  We examine the uncertainties in the pointing model 
applied to all SHARC-II data and conservatively find that the model 
corrections are good to within $\sim$3$''$, approximately $1/3$ of the SHARC-II 
beam.  We examine the differences between two array scan modes and find that 
the instrument calibration, beam size, and beam shape are  
similar between the two modes.  We also show that the same flux 
densities are measured when sources are observed in the two different modes, 
indicating that there are no systematic effects introduced into our catalog 
by utilizing two different scan patterns during the course of taking 
observations. We find a detection rate of 95\% for protostellar cores but only 
45\% for starless cores, and demonstrate the existence of a SHARC-II detection 
bias against all but the most massive and compact starless cores.  Finally, 
we discuss the improvements in protostellar classification enabled by 
these 350~\um\ observations.
\end{abstract}


\section{Introduction}
Stars form in dense cores of dust and molecular gas 
\citep[e.g.,][]{DiFrancesco2007,wardthompson2007:ppv}.  The ultraviolet, 
optical, and infrared radiation from both stars forming inside cores and 
the interstellar radiation field (ISRF) is absorbed by the dust within these 
cores, heating the dust to typical temperatures of $10-20$ K 
\citep[e.g.,][]{DiFrancesco2007}.  The dust then re-radiates this emission 
at submillimeter and millimeter wavelengths.  Thus, to study the very roots of 
star formation it is necessary to observe dense cores at these wavelengths.

Most nearby low-mass star-forming regions have been extensively surveyed at 
wavelengths between 850 \um\ and 1.3 mm due to the large number of bolometers 
available at these wavelengths and the high availability of weather suitable 
for $\sim$1 mm observations at most telescope sites \citep[e.g.,][]{Motte1998,Testi1998,Shirley2000,Johnstone2000,Johnstone2001,Motte2001a,Motte2001b,Young2003,kirk2005,1MMSurveyPers,Stanke2006,Young2006,1MMSurveyOph,1MMSurveySerp,Kauffmann2008}.  
Because they are both optically thin and in the Rayleigh-Jeans limit for 
$10-20$ K dust, continuum observations at $\sim$1 mm are ideally suited for 
tracing the total dust mass and easily pick out the dense star-forming cores \citep[e.g.,][]{1MMSurveySerp}.  However, since the peaks of 
$10-20$ K blackbodies occur at $250-500$ \um, these millimeter wavelength 
surveys do not constrain the peaks of the spectral energy distributions 
(SEDs).  Observations at shorter submillimeter wavelengths are needed to 
measure total source luminosities \citep{dunham2008:lowlum,dunham2013:luminosities,dunham2014:ppvi,enoch2009:protostars}, 
separate the contributions from internal 
(from a protostar) and external (from the ISRF) heating 
\citep{dunham2006:iram04191}, and accurately classify protostars into an 
evolutionary sequence \citep{andre1993:class0,chen1995:tbol,dunham2008:lowlum,dunham2014:ppvi,frimann2015:synthetic1}. 
Such observations are being provided by the {\it Herschel} Gould Belt Survey, 
which has obtained $70-500$ $\mu$m images of all of the nearby, star-forming 
clouds in the Gould Belt \citep[e.g.,][]{andre2010:hgbs}.  While this survey 
is providing unprecedented coverage and sensitivity of nearby star-forming 
regions at submillimeter wavelengths, it is doing so at relatively low 
angular resolution (ranging from approximately 9$''$ at 70 \um\ to 36$''$ at 
500 \um).

In an effort to provide a submillimeter catalog of dense cores with high 
spatial resolution, we present in this paper 350 $\mu$m continuum observations 
of low-mass protostellar cores taken with the Submillimeter High Angular 
Resolution Camera II (SHARC-II) at the Caltech Submillimeter Observatory 
(CSO) on Mauna Kea, Hawaii.  SHARC-II was a 350 and 450 $\mu$m ``CCD-style'' 
bolometer array of 12 $\times$ 32 pixels giving an instantaneous field of view 
(FOV) of 2.59$'$ $\times$ 0.97$'$, with the pixels filling over 90\% of 
the focal plane and separated by their projected size on the sky 
of 4.85$''$ \citep{dowell2003:sharc}.  With good focus 
and pointing, the 350 $\mu$m beam has a full-width at half-maximum (FWHM) 
of 8.5$''$.  In practice the effective resolution at 350 $\mu$m is 10$''$ 
(see Section \ref{sec_datareduction}), 
providing 2.5 times higher angular resolution 
than the {\it Herschel} Gould Belt survey at the same wavelength (albeit at 
lower sensitivity, especially to extended emission).  Targets were selected to 
provide complementary data to several large surveys of nearby, low-mass 
star-forming regions, including the {\it Spitzer Space Telescope} c2d 
\citep{evans2003:c2d,evans2009:c2d} and Gould Belt 
\citep{gutermuth2008:serpsouth,dunham2015:gb} Legacy 
Surveys and the {\it Herschel Space Observatory} DIGIT \citep[e.g.,][]{sturm2010:digit,vankempen2010:digit,green2013:digit,green2016:cdf} 
open time Key Project.  Some of the observations presented here were originally 
published by \citet{Wu2007}, but we have re-analyzed and included them here 
using an updated version of the data reduction software and improved pointing 
model corrections.  In general, we present observations of regions that are 
already well documented at other wavelengths so that this catalog will complement existing data in studies of the characteristics of protostellar 
regions. Though this is the primary goal of the paper, we also use our data to 
examine the sensitivity of SHARC-II to extended emission and to the observing 
mode used on the telescope, and to assess two different methods of classifying 
protostars.

We organize this paper as follows: First we describe the target selection and 
observation strategy in Sections \ref{sec_targets} and \ref{sec_observations}, 
respectively.  We then discuss the data reduction and calibration processes in 
Sections \ref{sec_datareduction} and \ref{sec_calibration}, respectively.  
We discuss the source extraction procedure in Section \ref{sec_extraction}, 
including an analysis of the effects of optimizing the data reduction pipeline 
for the recovery of extended emission.  We present our source catalog in 
Section \ref{sec_results}, including source positions, flux densities, 
and radii.  A comparison of the results from two different observing modes is 
presented in Section \ref{sec_mode}.  In Section \ref{sec_extended} we 
discuss the sensitivity of our observations to extended emission, and in 
Section \ref{sec_classification} we investigate the effects of including 
SHARC-II 350 \um\ photometry when classifying protostars.  
A summary of our results is presented in Section \ref{sec_summary}.

\section{Observations}

\subsection{Target Selection}\label{sec_targets}

Table 1 lists the targets of this survey, including the name of the core/cloud, 
the scan type (see below), the map center coordinates ordered by increasing 
Right Ascension, the distance to the target and reference for the distance 
determination, a representative reference for each target, 
the observation date, the 1$\sigma$ rms noise in units of 
mJy beam$^{-1}$, and the large cloud complex in which each target is located.  
As there is significant ambiguity in choosing a single representative 
reference for each target, we refer the reader to the SIMBAD 
database\footnote{Available at: http://simbad.u-strasbg.fr} for a 
comprehensive list of references for each object.  
The noise is measured as the standard deviation of all 
off-source pixels, calculated using the {\sc sky} procedure in the IDL 
Astronomy Library.  Two versions of each map are produced, one with and one 
without extended emission preserved (see Section 
\ref{sec_data_reduction_calibration} below for details); the noise is measured 
in the maps without extended emission preserved.  

As described in Section 1, the main purpose of this survey is to provide 
complementary 350 $\mu$m observations of low-mass star-forming clouds and 
cores observed in various {\it Spitzer} and {\it Herschel} survey programs.  
Thus targets were selected from the lists of regions included in those 
surveys, often focusing on individual studies that would benefit from these 
data.  As a result, the data presented here do not represent an unbiased 
submillimeter survey of star-forming regions, but a targeted survey designed 
to provide a catalog of useful complementary data.  

\subsection{Observations}\label{sec_observations}

Observations were conducted at the CSO in 
14 observing runs spread over seven years, ranging from May 2003 through 
December 2010.  All of the data obtained between May 2003 and November 2005 
were previously published (Wu et al.~2007); here we present updated 
images and catalogs using a newer version of the data reduction software 
(see Section 3).  These data were obtained using the sweep mode of SHARC-II, 
which moves the telescope following Lissajous curves in both the x and y 
dimensions.  This mode, which utilizes scan rates between 5--10 arcseconds 
s$^{-1}$ depending on the exact size mapped, results in a map 
with a fully sampled central region of uniform coverage, beyond which the 
coverage decreases and thus the noise increases.  The size of this central 
region depends on the exact observing parameters, but is typically $\sim$ 1$'$ 
$-$ 2$'$ for our observations.  Beginning in December 2006, we began 
experimenting with using the box-scan observing mode to map larger areas.  
This mode, which utilizes faster scan rates (typically in the range of 20--40 
arcseconds s$^{-1}$), moves the telescope in a straight line at a 45$^{\rm o}$ 
angle until it hits the boundary of a box, 
changes direction such that the angle of 
reflection equals the angle of incidence, and continues until the box is fully 
sampled.  The exact size of the box depends on the observing parameters, but 
is typically $\sim$ 6$'$ $-$ 10$'$ for our observations.  As the box-scan mode 
is optimized for mapping both larger areas and regions with extended emission, 
all data obtained during and after July 2008 were obtained exclusively in this 
mode.  Some sources were observed in both the Lissajous and box-scan observing 
modes.  In those cases, only the box-scan observations are listed in Table 1.  
The Lissajous observations for these sources will be discussed in Section 5, 
where we compare results from the two observing modes.

The total integration time was typically $30-120$ minutes per map, depending 
on weather conditions and the expected brightness of sources in each map.  
The noise levels of the final maps span nearly two orders of magnitude, ranging 
from approximately 8 to 500 mJy beam$^{-1}$ (see below).  Thus we caution that 
these maps form a very heterogeneous dataset in terms of sensitivity.  
Integrations for each map were broken into individual scans, each with a 
duration ranging from $5-15$ minutes depending on the stability of the 
atmosphere and the minimum time required to complete one scan in the chosen 
observing mode.  The zenith optical depth at 225 GHz ranged from $0.03-0.09$, 
with values of $\sim 0.05-0.07$ most typical.  With an approximate scaling 
factor of 20, these correspond to 350 $\mu$m zenith optical depths of 
$\sim 0.6-1.8$, with values of $\sim 1-1.4$ most typical.  During all of our 
observations except those obtained in June 2005 (see Wu et al.~2007 for more 
details), the Dish Surface Optimization System 
(DSOS)\footnote{See http://www.cso.caltech.edu/dsos/DSOS\_MLeong.html} was 
used to correct the dish surface figure for gravitational deformations as the 
dish moves in elevation during observations.  The pointing and focus were both 
checked and updated every $1-2$ hr each night, primarily with the planets 
Mars, Uranus, and Neptune, but occasionally with other secondary calibrators 
chosen from the SHARC-II website\footnote{See http://www.submm.caltech.edu/$\sim$sharc/analysis/calibration.htm}.  The pointing was further corrected in 
reduction based on a pointing model (see Section 3).

\section{Data Reduction and Calibration}\label{sec_data_reduction_calibration}

\subsection{Data Reduction}\label{sec_datareduction}

The data were reduced using the Comprehensive Reduction Utility for SHARC-II (CRUSH) version 2.12-1, a publicly available\footnote{See http://www.submm.caltech.edu/$\sim$sharc/crush/index.htm}, Java-based software package that iteratively solves for both the source signal and the various correlated noise components \citep[e.g.,][]{Kovaks2008a, Kovaks2008b}. For increased redundancy, we 
use CRUSH to add together the bolometer time streams from individual scans 
before obtaining the solutions, taking into account various noise elements 
and changing atmospheric conditions for each scan.  The only exception to 
this general method is NGC1333, which had to be broken into five smaller pieces first due to computer memory limitations. These five pieces were then coadded together using the {\sc coadd} function of the CRUSH package, after which point they were handled exactly the same way as the other sources. As described in 
more detail in Section \ref{sec_extraction} below, two versions of each map 
were produced: one with the extended flag given to CRUSH to optimize the software 
pipeline for the recovery of extended emission, and one without it given.  
The differences between and uses of the two versions of each map are discussed 
below.
	
The atmospheric opacity during each scan was determined from an online 
database\footnote{http://www.submm.caltech.edu/$\sim$sharc/analysis/taufit.htm}
of measurements of the zenith optical depth at 225 GHz, $\tau_{\rm 225 GHz}$, 
obtained in ten minute intervals.  This database includes polynomial fits 
to $\tau_{\rm 225 GHz}$ versus time for each night, where the orders of the 
polynomials were treated as free parameters.  The orders range from 3 to 90 
over the full database, but are typically less than 20, with a mean value of 
13.  We visually inspected plots of $\tau_{\rm 225 GHz}$ and the resulting 
polynomial fits for each night to verify that the fits accurately trace the 
variations in $\tau_{\rm 225 GHz}$ throughout the night.  CRUSH uses these 
polynomial fits to calculate the optical depth at the time each observation scan was taken and uses this value to correct the signal. Pointing corrections were applied to each scan with CRUSH to correct for residual telescope pointing errors.  
These corrections were determined using a publicly available 
model\footnote{See http://www.submm.caltech.edu/$\sim$sharc/analysis/pmodel} fit to all pointing data from that observing run.  
We applied this model using the option to also correct for a random drift with time by evaluating model residuals for pointing scans taken within a few hours before and after each science scan. The final maps were generated with 1.5$''$ 
pixels.  A Gaussian smoothing function with a FWHM of 4 pixels was applied to 
each map to reduce pixelation artifacts, resulting in a final effective beam 
of approximately 10$''$. 

\begin{figure}
\epsscale{1.2}
\plotone{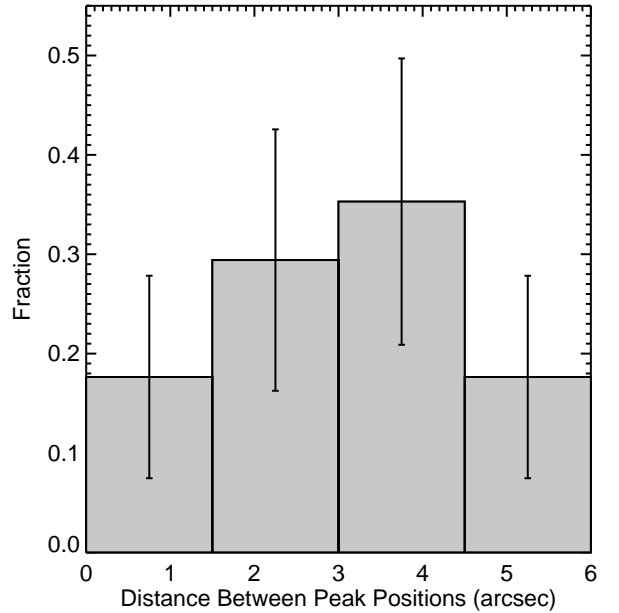}
\caption{\label{fig_boxlissdist}Histogram of the distances between measured 
peak positions for several sources that were observed twice on two
different dates, after applying the pointing model corrections.}
\end{figure}

\begin{figure}
\epsscale{1.2}
\plotone{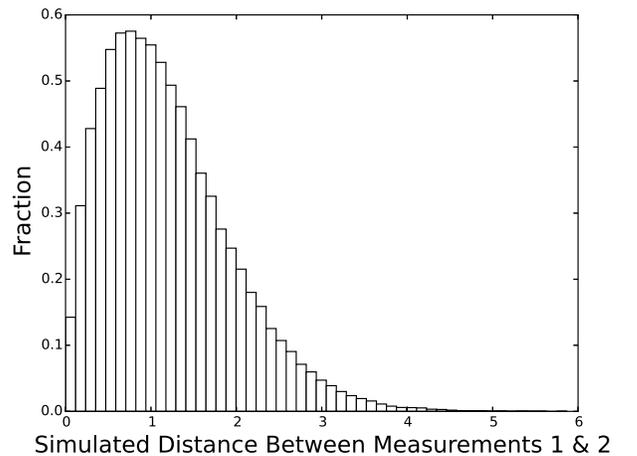}
\caption{\label{fig_montecarlo}Histogram showing the distance between measured 
peak positions for sources that were each observed twice in our Monte Carlo 
model, as explained in detail in the text.  The mean distance is 1.2 and the median distance is 1.1 (both in arbitrary units). Since the model assumed a pointing rms of 1, we find 
that the mean of this distribution is 20\% larger than the underlying rms of 
the pointing model.}
\end{figure}

In order to determine the residual pointing uncertainty after applying the 
pointing model corrections, Figure \ref{fig_boxlissdist} shows a histogram 
of the distances between measured peak positions for several sources that 
were observed twice on two different dates, after applying the pointing 
model corrections.  The mean distance is 2.9$''$.  In order to interpret this 
in terms of a residual rms uncertainty in the pointing model, we construct 
a Monte Carlo model in which a source is placed on a polar coordinate grid 
twice, with each angle drawn randomly from a uniform distribution and each 
radius drawn randomly from a Gaussian distribtuion with $\sigma = 1$.  Note 
that this $\sigma$ has no units as we are only concerned with obtaining a 
dimensionless ratio that will characterize the pointing model uncertainty, 
as described below.  The distribution of distances between the two 
``observations'' of each source in the Monte Carlo model is shown in Figure 
\ref{fig_montecarlo}; this distribution has a mean of 1.2.  Since the input to 
the simulation was a pointing model with an assumed rms of 1, and the resulting 
mean of the distance distribution is 20\% larger, we thus infer that our 
observed mean distance between peak positions of 2.9$''$ implies an underlying 
pointing model residual rms of 2.4$''$.  Given the small number of sources 
observed twice and uncertainties in the assumptions of the Monte Carlo model, 
we thus conservatively estimate that the pointing model corrections are good 
to within $\sim 3''$.

Once the maps were created using CRUSH, we used the {\sc imagetool} function 
of the CRUSH package to eliminate map edges with increased noise by removing 
all pixels in the map with less than 25\% of the maximum integration time. 
The average 1$\sigma$ rms for each map was calculated by using the {\sc sky} 
routine in the IDL Astronomy Library to measure the standard deviation of all 
off-source pixels and then calibrating with the peak calibration factor, as 
defined below. 

\subsection{Calibration}\label{sec_calibration}

While CRUSH adopts an approximate calibration factor to produce maps in 
calibrated units of Jy beam$^{-1}$, it does not account for the fact that the 
instrument calibration changes both with observing conditions and randomly 
with time\footnote{As described in the online CRUSH documentation for SHARC-II 
data reduction (available at: 
http://www.submm.caltech.edu/$\sim$sharc/crush/instruments/sharc2/), 
the exact calibration factor that CRUSH applies depends on the line-of-sight 
optical depth at 350 $\mu$m.  Other factors that affect the calibration 
factor but are not automatically accounted for by CRUSH include the detector 
temperature, optical configuration, cleanliness of the mirrors, focus quality, 
and DSOS status.}.  Furthermore, the presence of beam sidelobes mean that the 
flux of a point source measured in apertures of increasing size will also 
increase, whereas ideally the flux of a point source should be independent 
of aperture size.  We thus recalibrate all of our data as follows.

\begin{figure*}
\epsscale{1.0}
\plotone{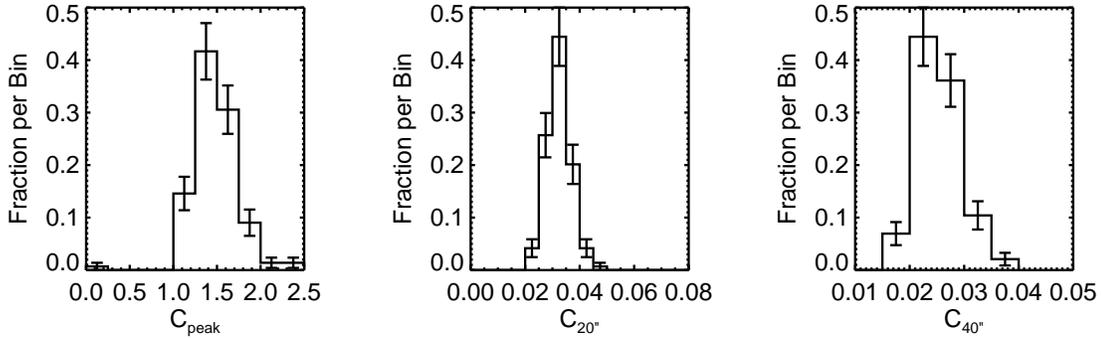}
\caption{\label{fig_histcal}Histograms for each set of calibration factors--peak, 20" aperture, and 40" aperture. The peak calibration factors are unitless. The 20" aperture and 40" aperture calibration factors convert from units of intensity (Jy beam$^{-1}$) to flux density (Jy) and therefore technically have units of Jy $/$ (Jy beam$^{-1}$) $=$ beam.}
\end{figure*}

\begin{figure*}
\epsscale{1.0}
\plotone{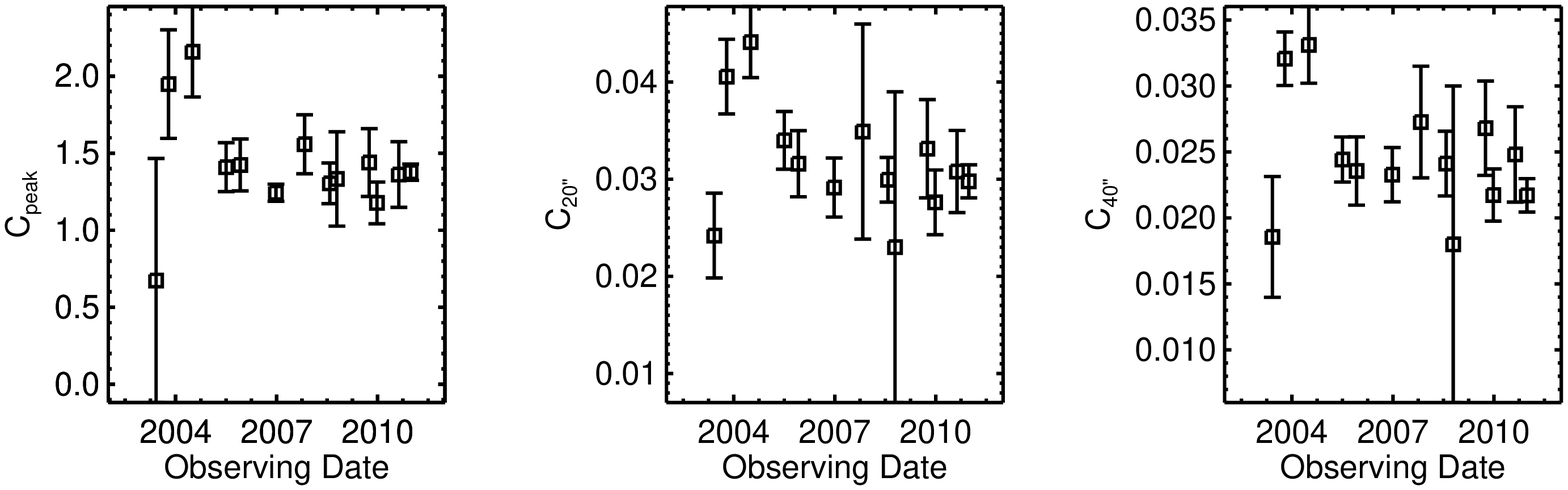}
\caption{\label{fig_calovertime}Mean calibration factors for each observing run from 2003 to 2010. The error bars are the standard deviation from each run. The first, second, and third panels are the beam, 20'' aperture, and 40'' aperture respectively. }
\end{figure*}

\begin{figure*}
\epsscale{1.0}
\plotone{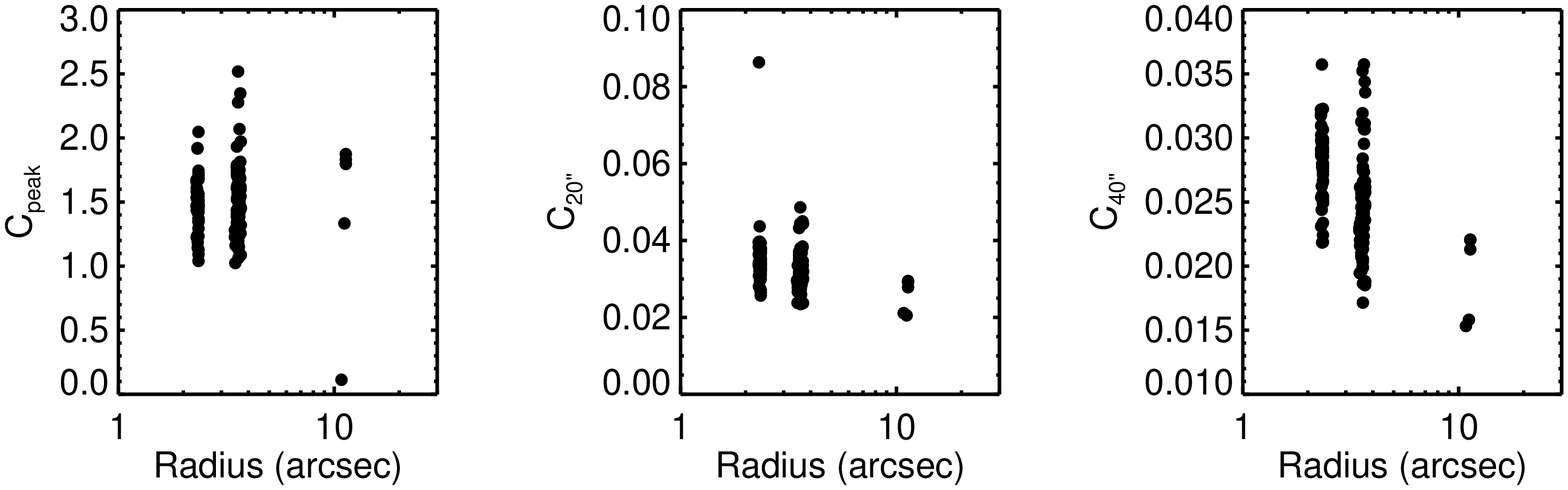}
\caption{\label{fig_caloverradius}Calibration factors from each planet scan plotted against the radius of the planet in that scan. The first, second, and third panels are the beam, 20'' aperture, and 40'' aperture respectively.}
\end{figure*}

%In each panel, the red trend line shows a horizontal linear regression while the blue trend line shows a sloped linear regression. In each panel, the reduced chi square value of the horizontal line is smaller than that of the sloped line, indicating that there is no trend in any panel.
%
%reduced chi squares-- 0.852, 0.858; 0.430, 0.434; 2.519, 2.555; 1.916, 1.937; 0.430, 0.488; 2.519, 3.249%

\begin{figure*}
\epsscale{1.0}
\plotone{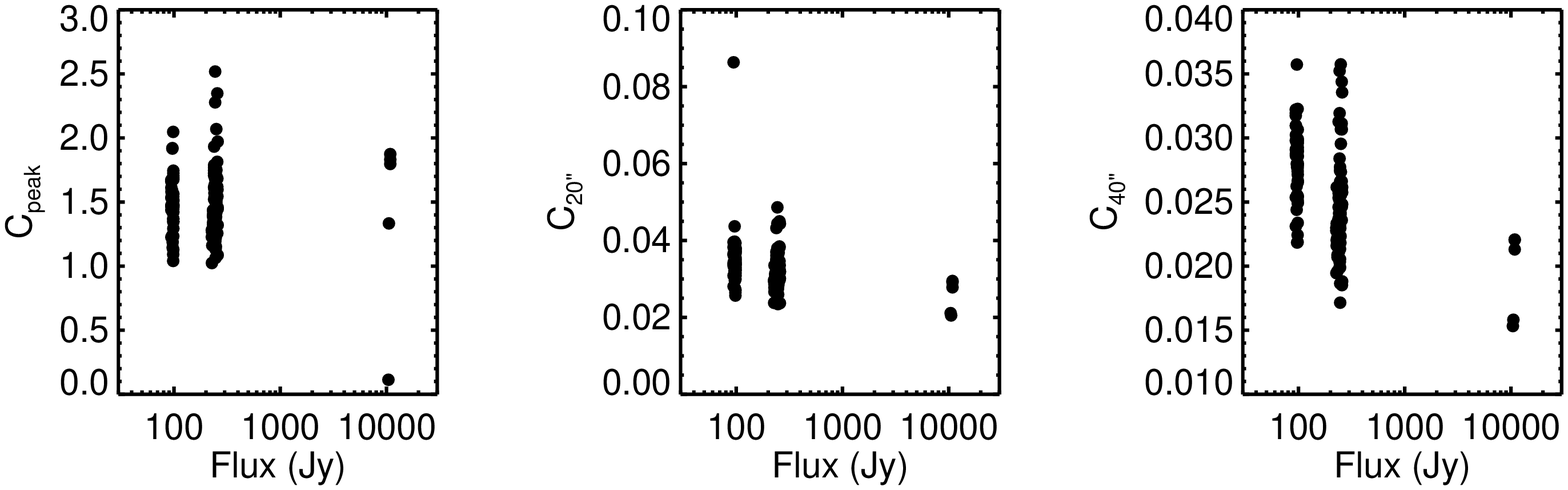}
\caption{\label{fig_caloverflux}Calibration factors from each planet scan plotted against the flux density of the planet in that scan. The first, second, and third panels are the beam, 20'' aperture, and 40'' aperture respectively. }
\end{figure*}

To derive calibration factors, we used observations of Mars, Neptune, and 
Uranus, which were observed every few hours to check and update the telescope 
pointing. These planet scans, which we hereafter refer to as calibration scans, 
were reduced with CRUSH and used to calculate the calibration factors. All 
calibration scans were observed using the Lissajous observing mode on the 
telescope, regardless of what observing mode was used for the science sources 
to which these calibrations were applied. As shown in Section \ref{sec_mode}, 
the calibration factors do not depend on observing mode, validating this 
strategy.  For each calibration scan, we measured the peak intensity and 
flux densities in 20$''$ and 40$''$ diameter apertures using custom IDL 
routines. We choose these aperture sizes to match previous (sub)millimeter 
continuum surveys that measure flux densities in standard apertures of 20$''$, 
40$''$, 80$''$, and 120$''$ in diameter \citep[e.g.,][]{1MMSurveyPers,1MMSurveyOph,Young2006,1MMSurveySerp,Wu2007,Kauffmann2008}.
Here we only adopt the two smallest apertures since our calibration images are 
too small to use larger apertures. By comparing the measured flux densities to 
the known fluxes of these planets, we calculate three calibration factors: 
C$_{\rm peak}$, C$_{\rm 20"}$, and C$_{\rm 40"}$.  C$_{\rm peak}$ is simply the 
factor required to obtain calibrated maps in units of Jy beam$^{-1}$.  Since 
the maps are already calibrated with an approximate calibration factor, 
the values of C$_{\rm peak}$ are unitless and are generally close to unity, as 
seen below.  C$_{\rm 20"}$ and C$_{\rm 40"}$ are 
``aperture calibration factors'' and have units of 
Jy~/~(Jy~beam$^{-1}$)~$=$~beam.  Multiplying the flux densities measured 
through aperture photometry by the aperture calibration factors for the 
same size aperture will give the flux density in that aperture in Jy.  
As mentioned above, this method corrects for the beam sidelobes such that the 
measured flux density of a point source is independent of aperture size 
\citep[e.g.,][]{Shirley2000, 1MMSurveyPers}.  

\begin{figure*}
\epsscale{1.0}
\plotone{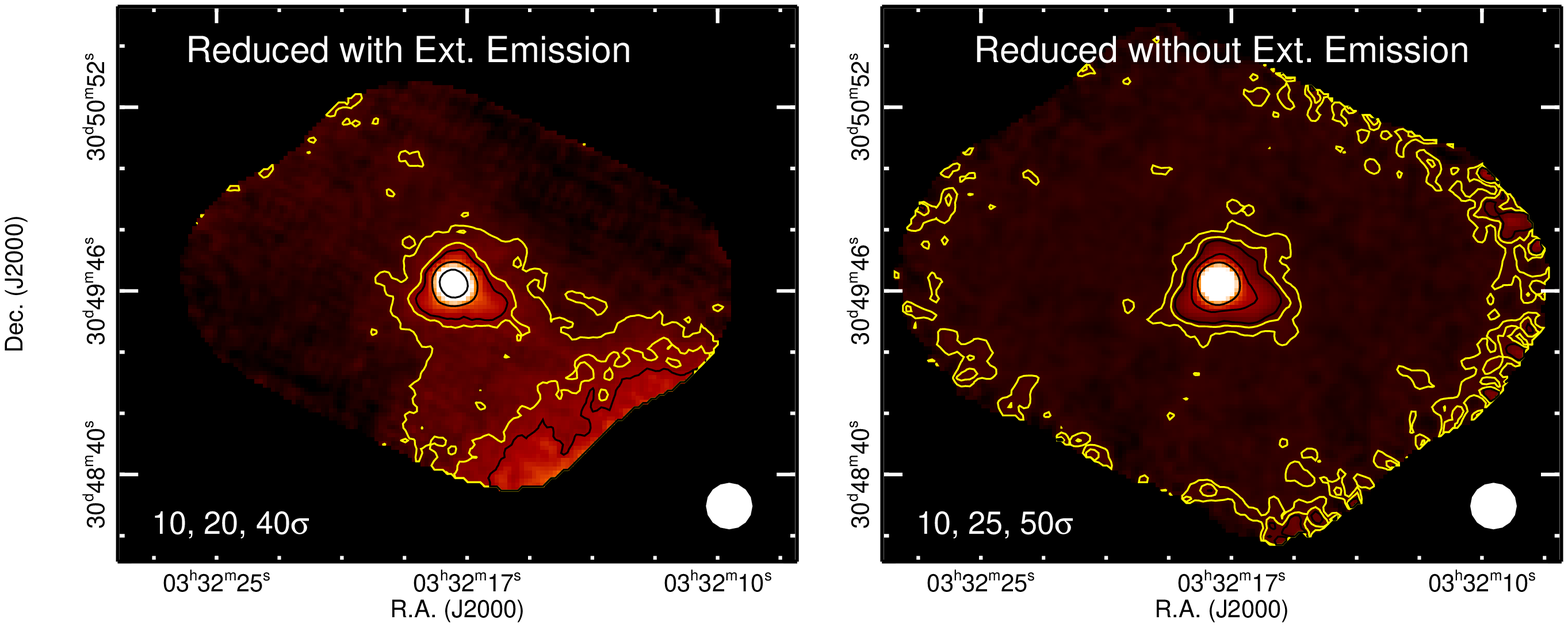}
\caption{IRAS 03292+3039 reduced with (left) and without (right) the extended 
emission flag included, with both panels displayed with a linear intensity 
color scale ranging from $-$0.4~to~3.0~Jy~beam$^{-1}$ (see Figure 
\ref{fig_scalebar} for a normalized version of the color scale bar used). 
The two yellow contours in each panel are plotted at 3$\sigma$ and 7$\sigma$, 
where 1$\sigma = 0.55$~Jy~beam$^{-1}$ in the map with the extended emission 
flag and 1$\sigma = 0.15$~Jy~beam$^{-1}$ in the map without the extended 
emission flag.  Additional contours are plotted in black, with the levels 
chosen manually for optimal visual display and printed in white text in the 
lower left corner of each panel.  The peak intensities 
(total flux densities in 40$''$ diameter apertures) are measured to be 
9.6~Jy~beam$^{-1}$ (17.4~Jy) in the map with the extended emission flag 
and 10.0~Jy~beam$^{-1}$ (14.1~Jy) in the map without the extended emission 
flag.}
\label{fig_wandwcompare}
\end{figure*}

In practice, C$_{\rm peak}$, C$_{\rm 20"}$, and C$_{\rm 40"}$ are calculated as 
follows.  For C$_{\rm peak}$, the expected peak intensity from the planet was 
calculated by convolving the SHARC-II beam (assumed to be Gaussian) with a 
disk of uniform brightness, using the known size and flux density of each 
planet on that observation date.  The resulting values are then divided by the 
measured peak intensity in each calibration scan.  For C$_{\rm 20"}$ and 
C$_{\rm 40"}$, the known total flux density of the planet on each observation 
date was divided by the measured flux density in 20$''$ and 40$''$ diameter 
apertures, respectively, using the uncalibrated maps.  The flux densities 
were measured using standard aperture photometry with no sky subtraction since 
CRUSH removes the background sky emission.  Table \ref{tab_calibrators1} 
lists our derived calibration factors for each observation night, and Table 
\ref{tab_calibrators2} lists the means and standard deviations of the 
calibration factors for each observing run.  Each map is calibrated using the 
mean calibration factors for that run; maps consisting of scans obtained 
over multiple runs are calibrated using the mean values over those runs.  
For maps observed in runs where no planet scans were taken, the mean 
calibration factors over all runs were used.

Figure \ref{fig_histcal} shows histograms for each of the calibration factors.  
To investigate whether the derived calibration factors vary with time or 
depend on the properties of the calibration source, Figures 
\ref{fig_calovertime}, \ref{fig_caloverradius}, and \ref{fig_caloverflux} 
plot the calibration factors versus observation date, angular size of the 
calibration source, and total flux of the calibration source.  No 
systematic variations are seen, and linear least squares fits to each panel 
in Figures \ref{fig_caloverradius} and \ref{fig_caloverflux} give better 
fits (lower reduced $\chi^2$ values) for zeroth order polynomial than for 
first order polynomials, indicating that any dependences of the calibration 
factors on source properties are smaller than the overall calibration 
uncertainties.  We calculate these calibration uncertainties by dividing the 
standard deviations by the means, resulting in values of 18\% for 
C$_{\rm peak}$, 21\% for C$_{\rm 20"}$, and 17\% for C$_{\rm 40"}$.  We thus 
conservatively adopt an overall calibration uncertainty of 25\%.
	 
\subsection{Source Extraction}\label{sec_extraction}

\begin{figure*}
\epsscale{1.0}
\plotone{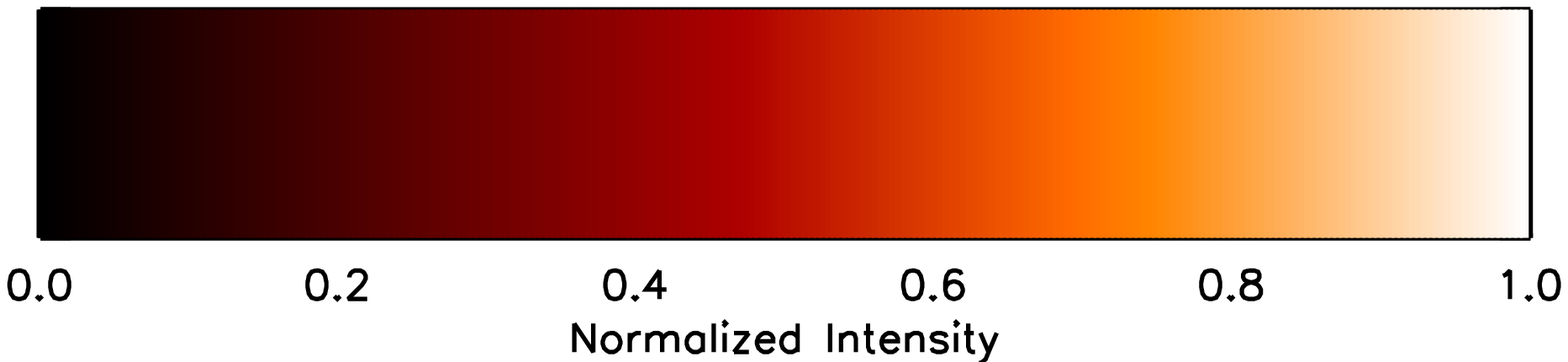}
\caption{Normalized intensity color scale bar for all of the images presented 
in this paper (Figures \ref{fig_wandwcompare}, 
\ref{fig_L1448}--\ref{fig_multiple12}, and 
\ref{fig_lissmultiple1}--\ref{fig_lissmultiple4}).  The minimum (black) and 
maximum (white) intensities for each map are presented in their respective 
figure captions.}
\label{fig_scalebar}
\end{figure*}

\begin{figure*}
\epsscale{1.0}
\plotone{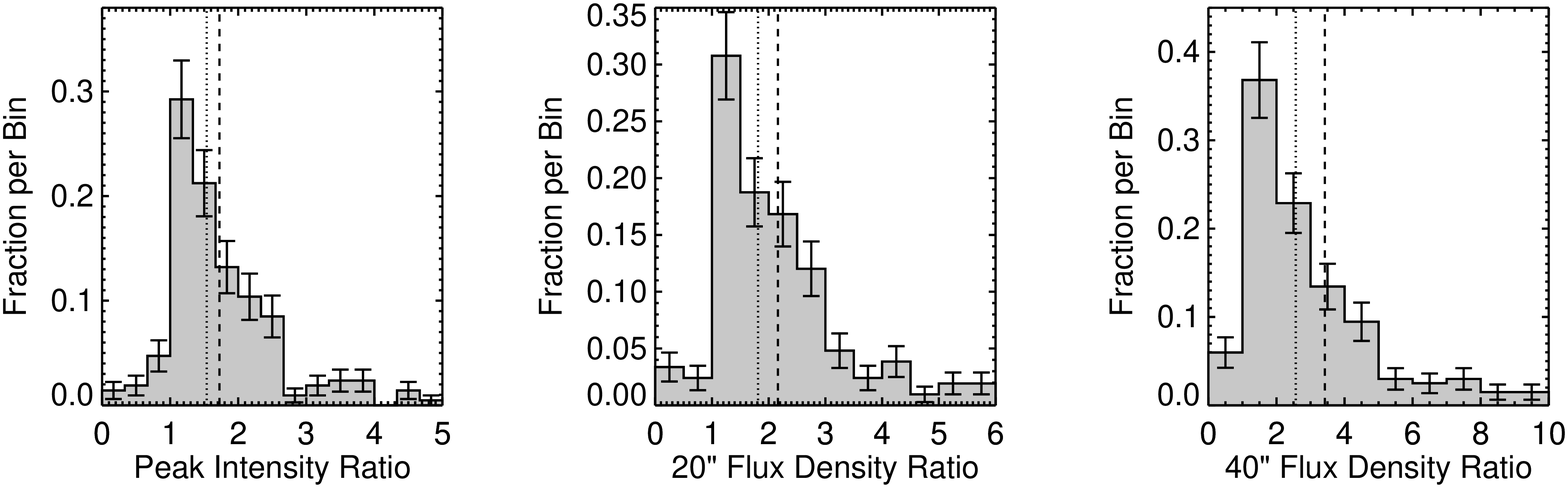}
\caption{Histograms for the ratios of flux densities measured in maps with 
the extended flag divided by those measured in maps without the extended flag, 
for each aperture size:  The peak (beam) intensities (left), the flux 
densities in 20$''$ diameter apertures (middle), and the flux densities in 
40$''$ diameter apertures (right).  The vertical dashed lines in each panel 
shows the mean of each distribution, and the vertical dotted lines show the 
medians.  The error bars on each bin are the statistical ($\sqrt{N}$) 
uncertainties.}
\label{fig_histwandw}
\end{figure*}

As noted above, two versions of each map were produced, one with and one 
without the extended flag given to CRUSH.  This flag optimizes the CRUSH 
pipeline for extended sources and results in maps that preserve extended 
emission at the cost of increased noise \citep[see][for details]{Kovaks2008a}.  
Since our targets are star-forming regions expected to feature extended 
emission, we investigated the possibility of using this flag to recover more 
extended emission.  However, we found that the added noise is dominated by sky 
noise that is temporally correlated in the time stream and thus spatially 
correlated in the final maps, rather than random, and thus especially 
problematic for source extraction.  Many false sources are detected and 
extracted regardless of the detailed implementation of source extraction.  
Figure \ref{fig_wandwcompare} shows an example of a map reduced with and 
without the extended flag.  Similar color figures are presented for all 
of the regions mapped below.  All maps are displayed with a linear intensity 
scaling with the minimum (black) and maximum (white) intensities given in 
their respective figure captions.  Figure \ref{fig_scalebar} displays a 
normalized intensity color scale bar that, combined with the minimum and 
maximum intensities given in the figure captions, can be used to determine 
the absolute intensities of each map.

Figure \ref{fig_wandwcompare} illustrates that the extended flag results in 
large areas of correlated noise that are extracted as sources.  Thus, in order 
to ensure a reliable catalog of sources, we perform source extraction on 
the maps produced without the extended flag (such as the one shown in the 
right panel of Figure \ref{fig_wandwcompare}).  
We extracted sources from each map using the Bolocat\footnote{Available at https://github.com/low-sky/idl-low-sky/tree/master/bolocat} source extraction 
routine \citep{rosolowsky2010:bolocat}.  Bolocat works by identifying regions 
of statistically significant emission based on their significance with respect 
to a local estimate of the noise in the maps.  These regions of high 
significance are then subdivided into multiple sources based on the presence 
of local maxima within the originally defined regions, with each pixel assigned 
to one of the sources using a seeded watershed algorithm, similar to the 
Clumpfind or Source EXtractor algorithms 
\citep{williams1994:clumpfind,bertin1996}.  Bolocat was previously used to 
extract sources from SHARC-II images of massive star-forming clumps by 
\citet{merello2015}, demonstrating the feasibility of using this source 
extraction routine on data from the SHARC-II instrument.

Bolocat requires three input parameters, all of which are measured in units 
of the map rms:  $P_{\rm amp}$, the minimum required amplitude for a source 
to be extracted; $P_{\rm base}$, the base level of emission out to which 
the initial detected source is expanded; and $P_{\rm deb}$, a source 
deblending parameter.  In practice, Bolocat first masks all regions of the map 
below $P_{\rm amp}$ and extracts one or more initial sources based on the 
number of regions of contiguous pixels remaining in the masked map.  It then 
expands each of these initial sources to encompass all contiguous pixels down 
to the level $P_{\rm base}$, with the rationale being that marginally 
significant emission levels spatially connected to those of higher 
significance are likely to be real.  Finally, each initial source is deblended 
into one or more sources by finding pairs of local maxima and identifying 
them as separate sources if the lower of the two is at least $P_{\rm deb}$ 
above the highest contour level that contains both maxima 
\citep[see][for details]{rosolowsky2010:bolocat}.  We adopted values of 
$P_{\rm amp} = 3$, $P_{\rm base} = 3$, and $P_{\rm deb} = 2$ based on matching 
``by-eye'' extractions in an initial exploration of the parameter space (note 
that, since $P_{\rm amp} = P_{\rm base}$, we did not expand the sources beyond 
their initial detections).  While \citet{merello2015} also adopted 
$P_{\rm amp} = 3$, they adopted lower values of the other two parameters 
($P_{\rm base} = 1$ and $P_{\rm deb} = 1$).  We found that 
the adoption of such low values for $P_{\rm base}$ and $P_{\rm deb}$ resulted in 
extractions that did not match either our best ``by-eye'' results or known 
sources from catalogs at other wavelengths.  In particular, the lower value of 
$P_{\rm deb}$ resulted in several objects that were clearly single being broken 
into multiple objects due to small noise variations.  The advantage of studying 
known regions with copious multiwavelength data in the literature, as opposed 
to \citet{merello2015}, allowed us to refine the best values of the source 
extraction parameters.

After extracting sources and measuring their deconvolved radii using the maps 
reduced without the extended flag, aperture photometry was performed on the 
maps produced with the extended flag, at the peak position of each extracted 
source.  We used custom IDL routines to measure the peak intensities and flux 
densities in 20$''$ and 40$''$ diameter apertures.  In cases of overlapping 
apertures with nearby sources, only those up to the largest in which 
overlap did not occur were kept.  We measured the uncertainties in the flux 
densities in each aperture by adding in quadrature the statistical uncertainty 
from the measurement itself and the overall calibration uncertainty of 25\%.

This method of using both sets of maps ensures that 
a reliable catalog of sources 
is extracted while also preserving as much extended emission as possible in 
the measured flux densities.  Figure \ref{fig_histwandw} shows the 
distribution of ratios of flux densities measured in the maps with the 
extended flag to those measured in the maps without the extended flag, for the 
peak intensities as well as the flux densities measured in 20$''$ and 40$''$ 
diameter apertures.  The mean (median) ratios are 1.7 (1.5), 2.2 (1.8), and 
3.4 (2.6) for the peak, 20$''$, and 40$''$ measurements.  
Thus, as expected, we see that the extended flag 
improves the flux recovery, especially in the larger apertures that are more 
sensitive to the amount of extended emission.  

\begin{figure*}
\epsscale{0.55}
\plotone{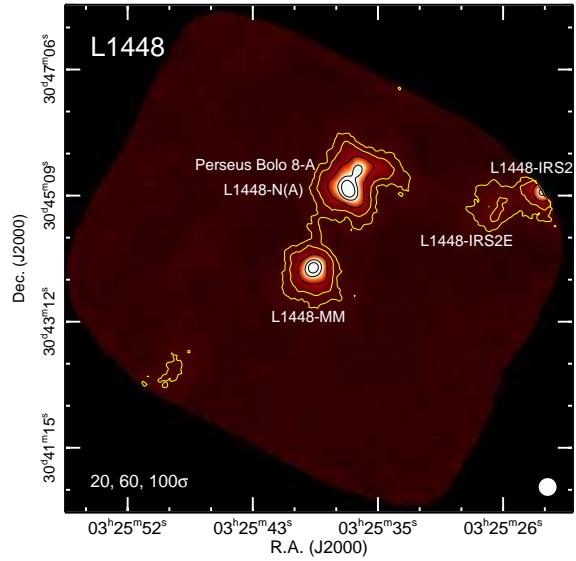}
\vspace{-0.3in}
\caption{SHARC-II 350 $\mu$m maps of the regions listed in Table 1, reduced 
without the extended emission flag and displayed on a linear intensity color 
scale, here L1448.  Maps with multiple sources have each source labeled. The 
beam size is shown at the lower right of each map.  The two yellow contour 
levels are plotted at 3$\sigma$ and 7$\sigma$, where 
1$\sigma = 0.151$~Jy~beam$^{-1}$.  Additional contours are plotted in black, 
with the levels chosen manually for optimal visual display.  These levels are 
printed in white text at the bottom of each panel, with no text indicating 
no additional black contours are plotted.  Emission seen toward the edges of 
the maps is not reliable and should be ignored.  The color scaling uses a 
linear intensity scale ranging from $-$1.3~to~8.0~Jy~beam$^{-1}$ (see Figure 
\ref{fig_scalebar} for a normalized version of the adopted color scale bar).}
\label{fig_L1448}
\end{figure*}

\begin{figure*}
\epsscale{1.1}
\plotone{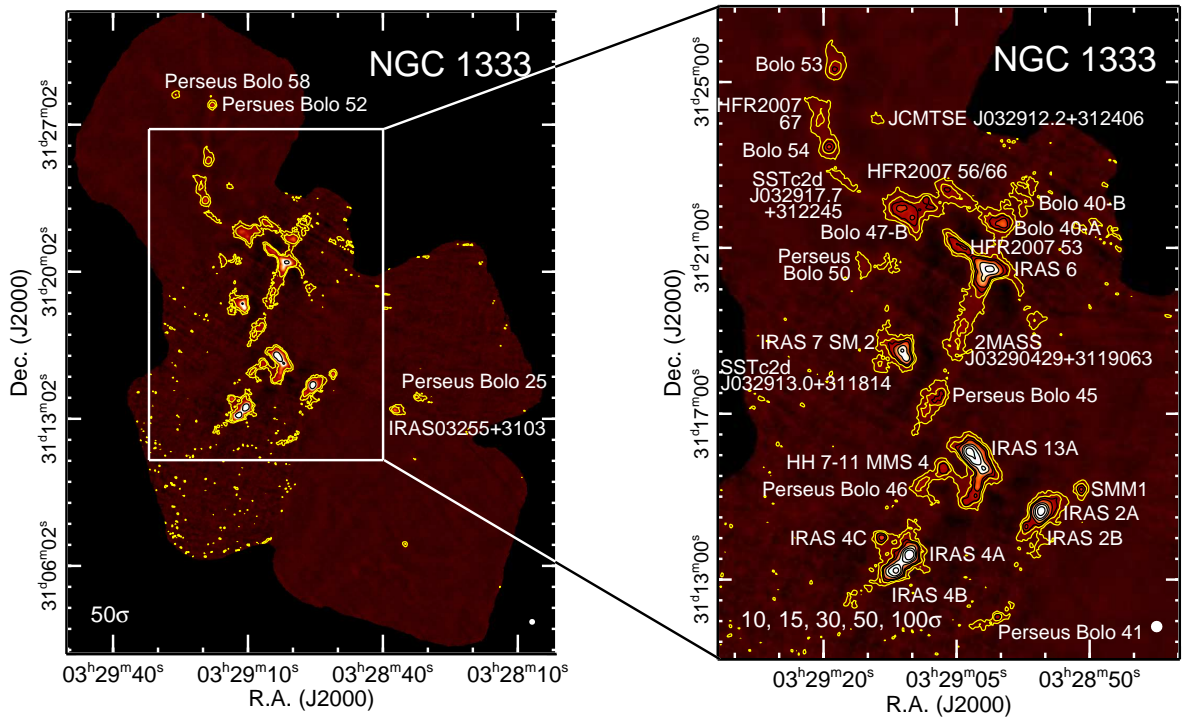}
\caption{Same as Figure \ref{fig_L1448}, except for 
NGC1333 ($1\sigma$~=~0.170~Jy~beam$^{-1}$, min~=~$-$1.2~Jy~beam$^{-1}$, max~=~5.8~Jy~beam$^{-1}$).  
A few sources in the right panel are labeled as Bolo XX rather than Perseus 
Bolo XX to save space.}
\label{fig_NGC1333}
\end{figure*}

\section{Results}\label{sec_results}

\begin{figure*}
\epsscale{0.95}
\plotone{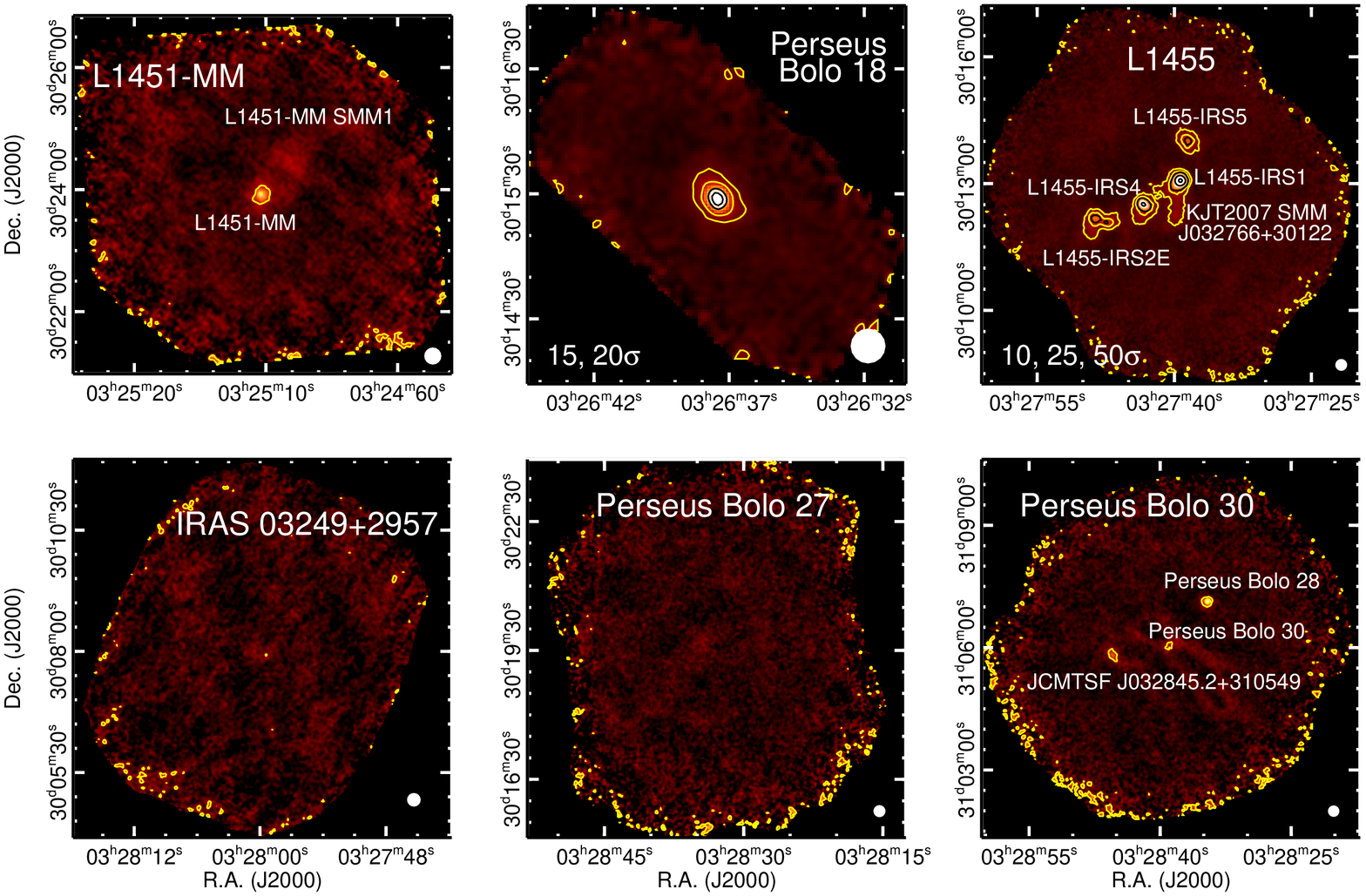}
\caption{Same as Figure \ref{fig_L1448}, except for 
L1451-MM ($1\sigma$~=~0.089~Jy~beam$^{-1}$, min~=~$-$0.1~Jy~beam$^{-1}$, max~=~0.7~Jy~beam$^{-1}$), 
Perseus Bolo 18 ($1\sigma$~=~0.104~Jy~beam$^{-1}$, min~=~$-$0.3~Jy~beam$^{-1}$, max~=~1.9~Jy~beam$^{-1}$), 
L1455 ($1\sigma$~=~0.067~Jy~beam$^{-1}$, min~=~$-$0.5~Jy~beam$^{-1}$, max~=~2.3~Jy~beam$^{-1}$), 
IRAS 03249+2957 ($1\sigma$~=~0.105~Jy~beam$^{-1}$, min~=~$-$0.1~Jy~beam$^{-1}$, max~=~0.9~Jy~beam$^{-1}$), 
Perseus Bolo 27 ($1\sigma$~=~0.104~Jy~beam$^{-1}$, min~=~$-$0.1~Jy~beam$^{-1}$, max~=~0.8~Jy~beam$^{-1}$), and 
Perseus Bolo 30 ($1\sigma$~=~0.122~Jy~beam$^{-1}$, min~=~$-$0.2~Jy~beam$^{-1}$, max~=~1.1~Jy~beam$^{-1}$).}
\label{fig_multiple1}
\end{figure*}

\begin{figure*}
\epsscale{0.95}
\plotone{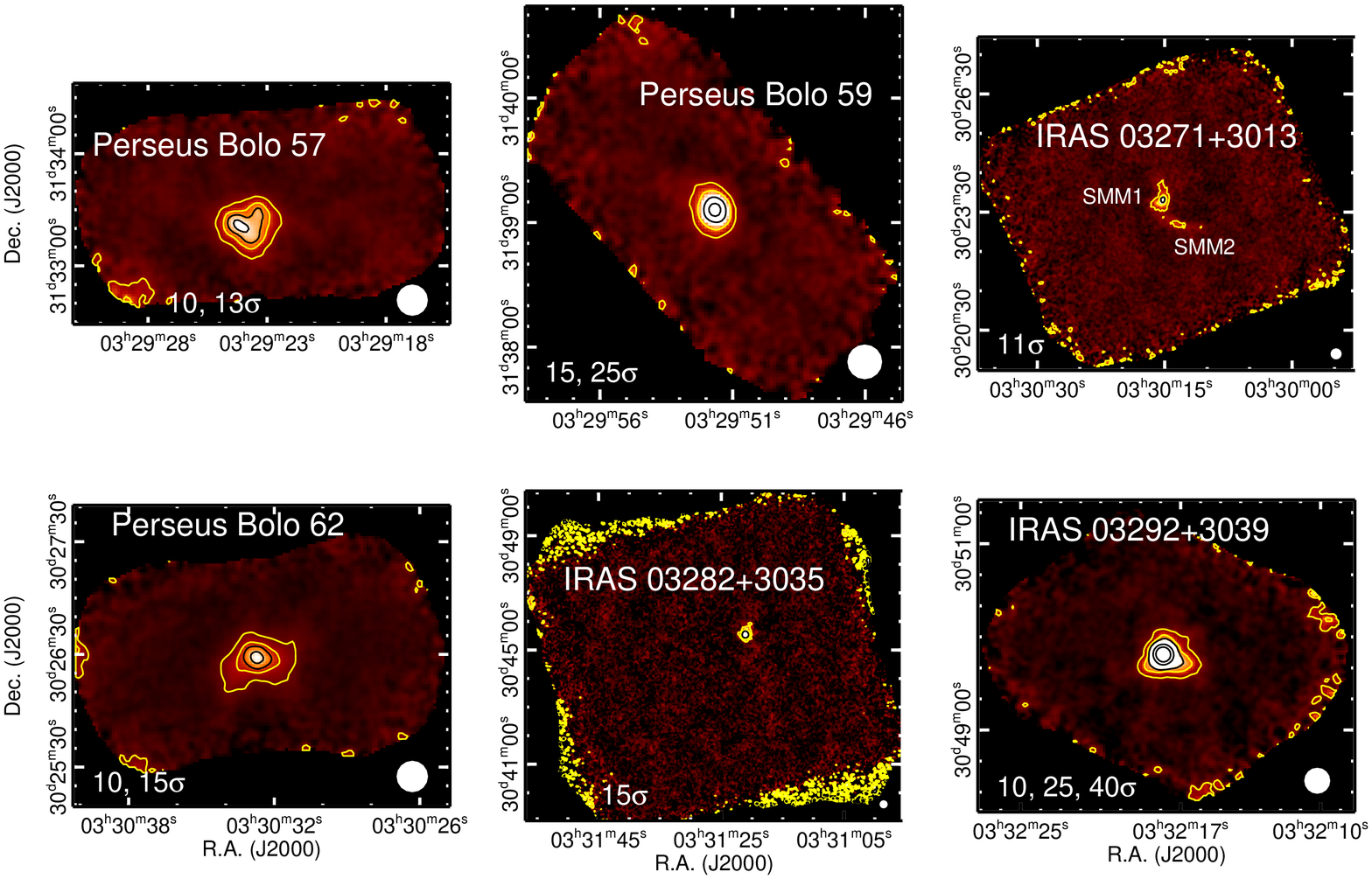}
\caption{Same as Figure \ref{fig_L1448}, except for 
Perseus Bolo 57 ($1\sigma$~=~0.091~Jy~beam$^{-1}$, min~=~$-$0.3~Jy~beam$^{-1}$, max~=~1.2~Jy~beam$^{-1}$), 
Perseus Bolo 59 ($1\sigma$~=~0.114~Jy~beam$^{-1}$, min~=~$-$0.3~Jy~beam$^{-1}$, max~=~1.2~Jy~beam$^{-1}$), 
IRAS 03271+3013 ($1\sigma$~=~0.099~Jy~beam$^{-1}$, min~=~$-$0.2~Jy~beam$^{-1}$, max~=~1.0~Jy~beam$^{-1}$), 
Perseus Bolo 62 ($1\sigma$~=~0.044~Jy~beam$^{-1}$, min~=~$-$0.1~Jy~beam$^{-1}$, max~=~0.7~Jy~beam$^{-1}$), 
IRAS 03282+3035 ($1\sigma$~=~0.174~Jy~beam$^{-1}$, min~=~$-$0.1~Jy~beam$^{-1}$, max~=~2.0~Jy~beam$^{-1}$), and 
IRAS 03292+3039 ($1\sigma$~=~0.152~Jy~beam$^{-1}$, min~=~$-$0.2~Jy~beam$^{-1}$, max~=~1.7~Jy~beam$^{-1}$).}
\label{fig_multiple2}
\end{figure*}

\begin{figure*}
\epsscale{0.95}
\plotone{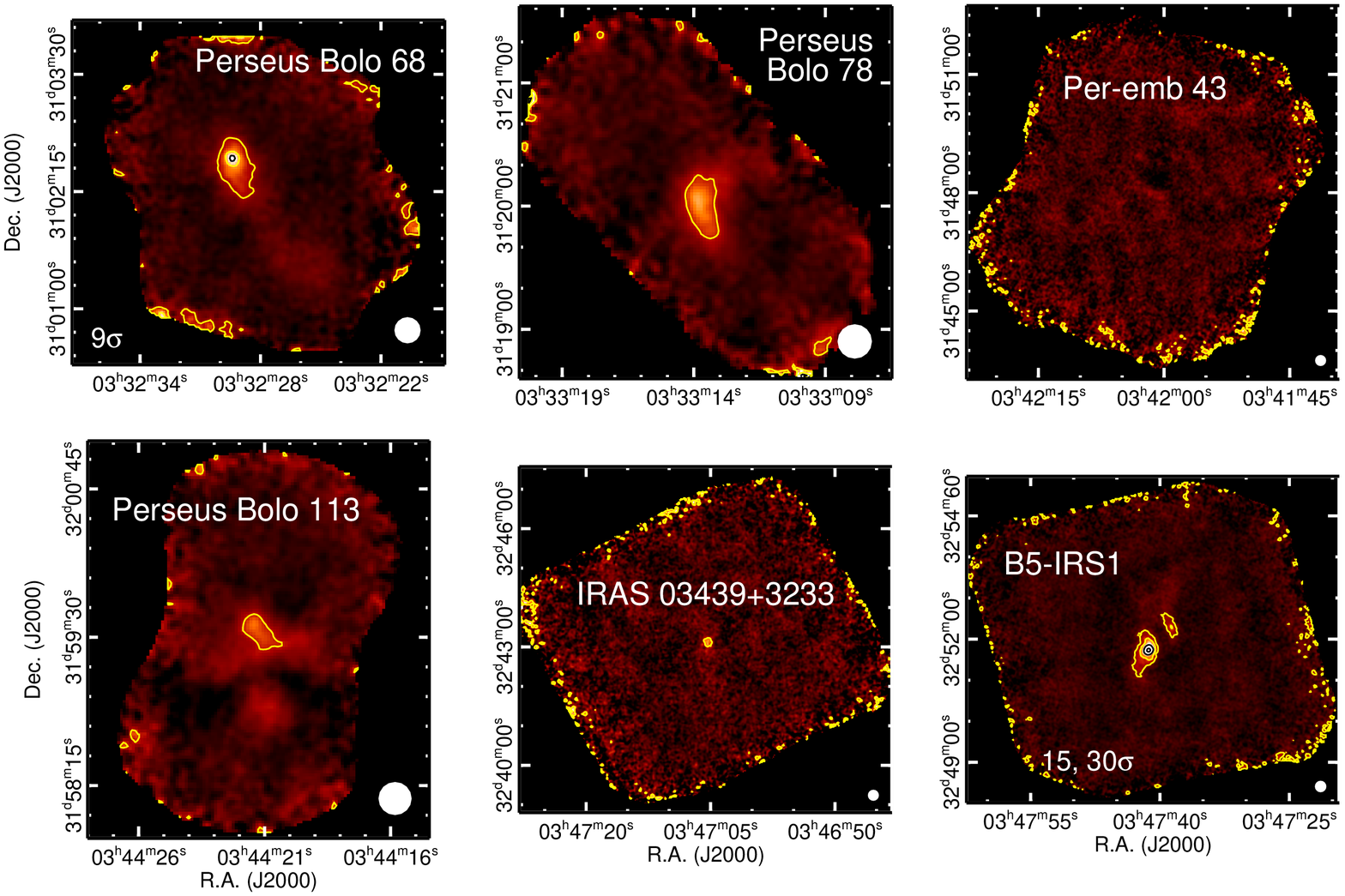}
\caption{Same as Figure \ref{fig_L1448}, except for 
Perseus Bolo 68 ($1\sigma$~=~0.096~Jy~beam$^{-1}$, min~=~$-$0.1~Jy~beam$^{-1}$, max~=~0.9~Jy~beam$^{-1}$), 
Perseus Bolo 78 ($1\sigma$~=~0.111~Jy~beam$^{-1}$, min~=~$-$0.2~Jy~beam$^{-1}$, max~=~0.8~Jy~beam$^{-1}$), 
Per-emb 43 ($1\sigma$~=~0.083~Jy~beam$^{-1}$, min~=~$-$0.1~Jy~beam$^{-1}$, max~=~0.6~Jy~beam$^{-1}$), 
Perseus Bolo 113 ($1\sigma$~=~0.087~Jy~beam$^{-1}$, min~=~$-$0.1~Jy~beam$^{-1}$, max~=~0.6~Jy~beam$^{-1}$), 
IRAS 03439+3233 ($1\sigma$~=~0.176~Jy~beam$^{-1}$, min~=~$-$0.1~Jy~beam$^{-1}$, max~=~0.5~Jy~beam$^{-1}$), 
and B5-IRS1 ($1\sigma$~=~0.095~Jy~beam$^{-1}$, min~=~$-$0.2~Jy~beam$^{-1}$, max~=~1.2~Jy~beam$^{-1}$).}
\label{fig_multiple3}
\end{figure*}

\begin{figure*}
\epsscale{0.7}
\plotone{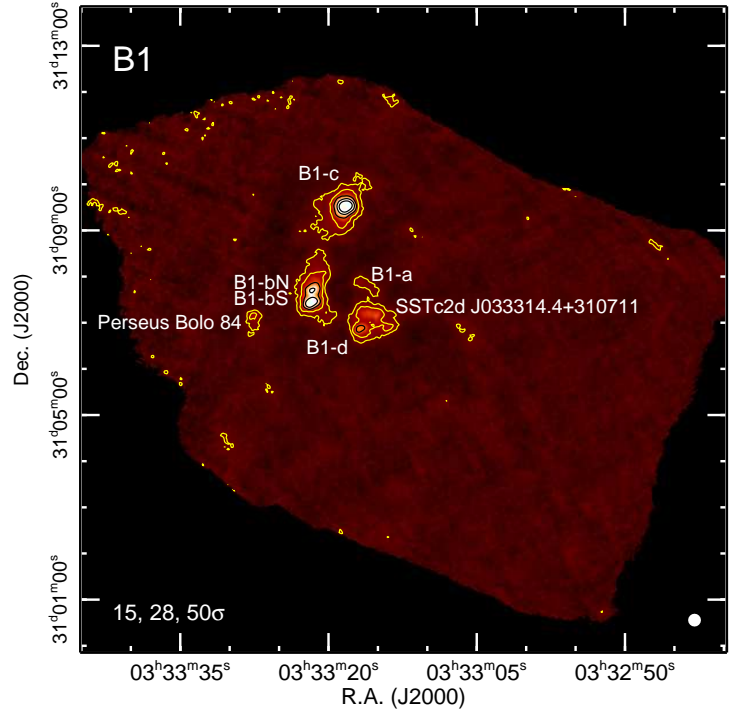}
\caption{Same as Figure \ref{fig_L1448}, except for 
B1 ($1\sigma$~=~0.121~Jy~beam$^{-1}$, min~=~$-$0.9~Jy~beam$^{-1}$, max~=~3.7~Jy~beam$^{-1}$).}
\label{fig_B1}
\end{figure*}

\begin{figure*}
\epsscale{1.0}
\plotone{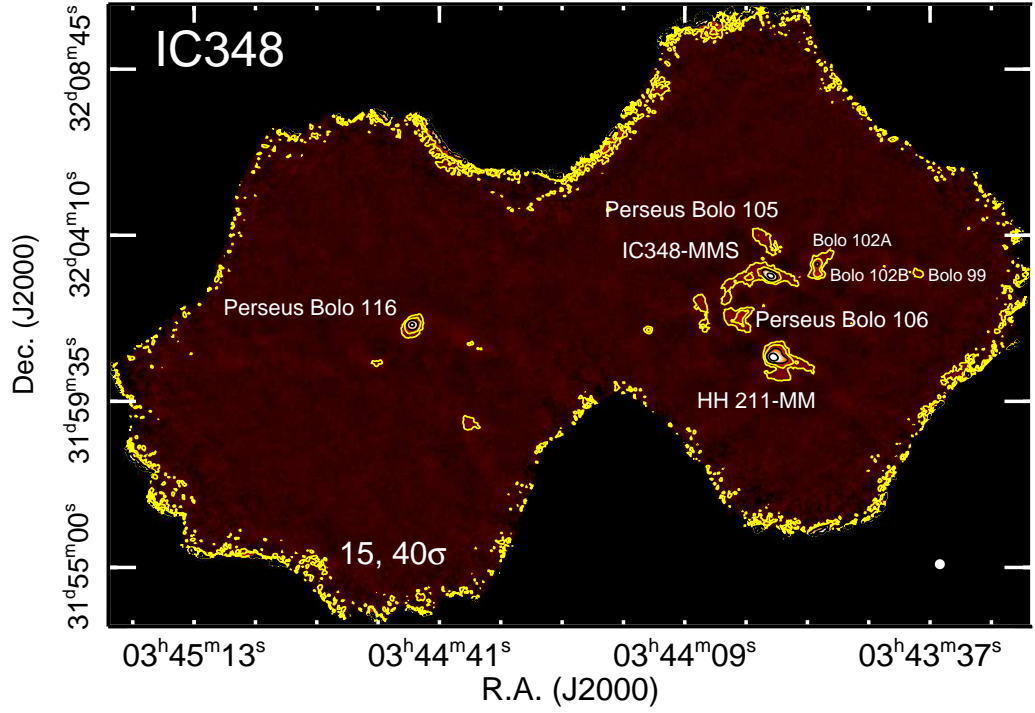}
\caption{Same as Figure \ref{fig_L1448}, except for 
IC348 ($1\sigma$~=~0.121~Jy~beam$^{-1}$, min~=~$-$0.6~Jy~beam$^{-1}$, max~=~3.3~Jy~beam$^{-1}$).}
\label{fig_IC348}
\end{figure*}

\begin{figure*}
\epsscale{0.95}
\plotone{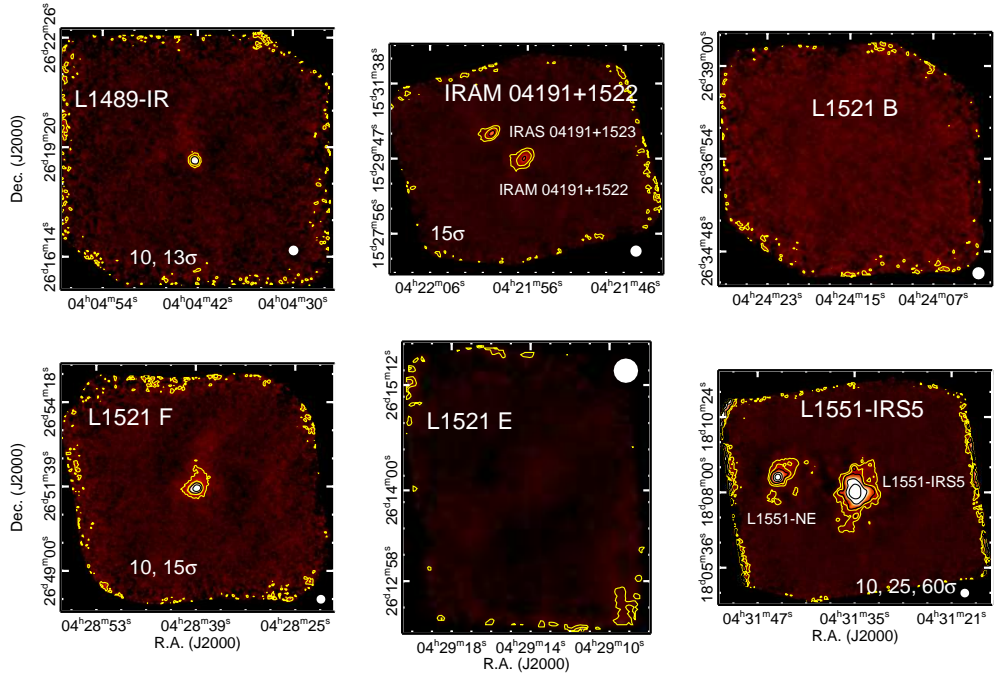}
\caption{Same as Figure \ref{fig_L1448}, except for 
L1489-IR ($1\sigma$~=~0.153~Jy~beam$^{-1}$, min~=~$-$0.3~Jy~beam$^{-1}$, max~=~0.2~Jy~beam$^{-1}$), 
IRAM 04191+1522 ($1\sigma$~=~0.080~Jy~beam$^{-1}$, min~=~$-$0.2~Jy~beam$^{-1}$, max~=~2.3~Jy~beam$^{-1}$), 
L1521B ($1\sigma$~=~0.091~Jy~beam$^{-1}$, min~=~$-$0.3~Jy~beam$^{-1}$, max~=~1.3~Jy~beam$^{-1}$), 
L1521F ($1\sigma$~=~0.056~Jy~beam$^{-1}$, min~=~$-$0.1~Jy~beam$^{-1}$, max~=~0.7~Jy~beam$^{-1}$), 
L1521E ($1\sigma$~=~0.087~Jy~beam$^{-1}$, min~=~0.1~Jy~beam$^{-1}$, max~=~2.1~Jy~beam$^{-1}$), 
and L1551-IRS5 ($1\sigma$~=~0.123~Jy~beam$^{-1}$, min~=~$-$0.4~Jy~beam$^{-1}$, max~=~2.7~Jy~beam$^{-1}$).}
\label{fig_multiple4}
\end{figure*}

\begin{figure*}
\epsscale{0.95}
\plotone{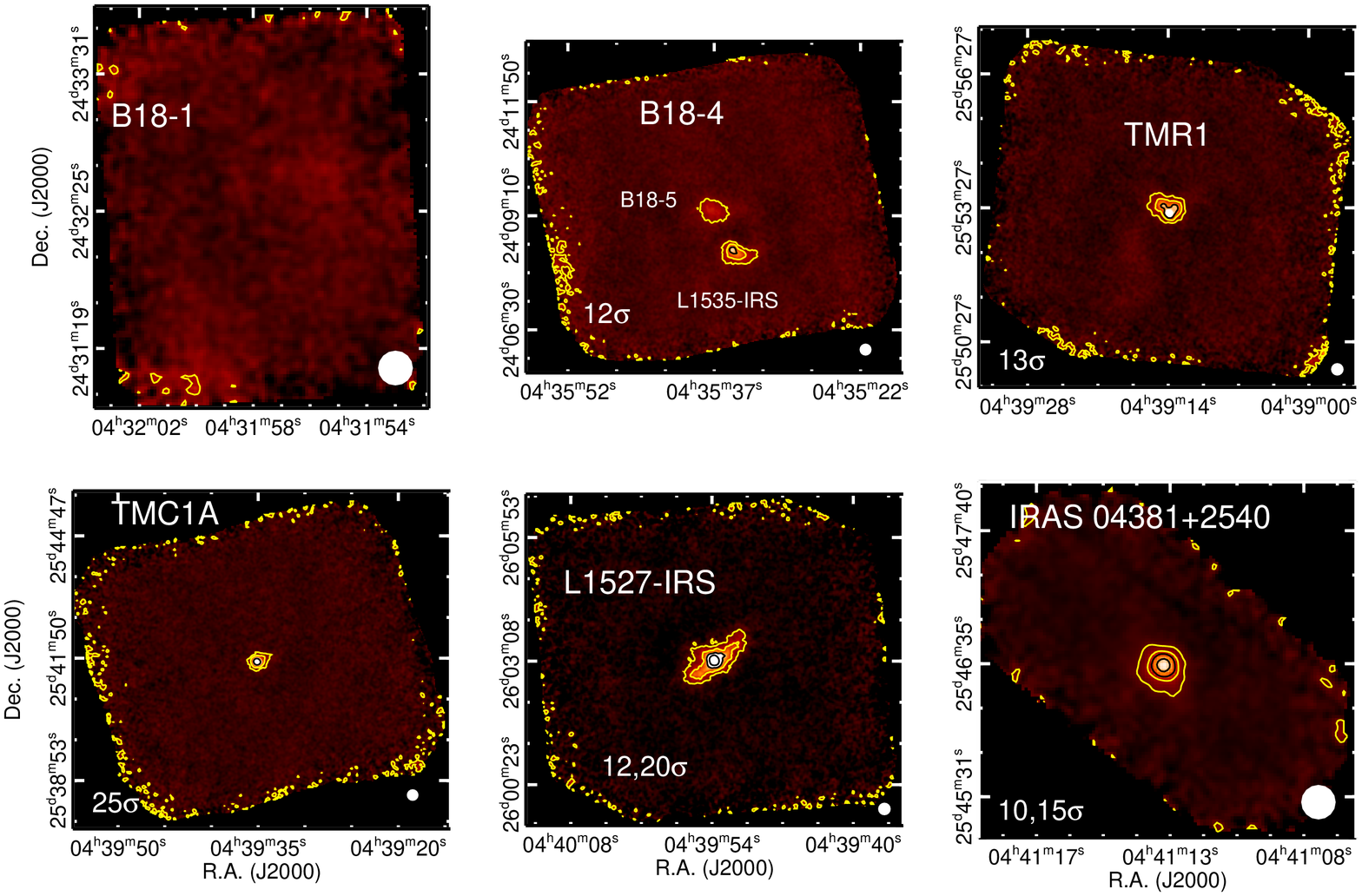}
\caption{Same as Figure \ref{fig_L1448}, except for 
B18-1 ($1\sigma$~=~0.107~Jy~beam$^{-1}$, min~=~$-$0.3~Jy~beam$^{-1}$, max~=~1.1~Jy~beam$^{-1}$), 
B18-4 ($1\sigma$~=~0.073~Jy~beam$^{-1}$, min~=~$-$0.3~Jy~beam$^{-1}$, max~=~1.2~Jy~beam$^{-1}$), 
TMR1 ($1\sigma$~=~0.084~Jy~beam$^{-1}$, min~=~$-$0.3~Jy~beam$^{-1}$, max~=~1.5~Jy~beam$^{-1}$), 
TMC1A ($1\sigma$~=~0.114~Jy~beam$^{-1}$, min~=~$-$0.4~Jy~beam$^{-1}$, max~=~2.3~Jy~beam$^{-1}$), 
L1527-IRS ($1\sigma$~=~0.155~Jy~beam$^{-1}$, min~=~$-$0.1~Jy~beam$^{-1}$, max~=~2.1~Jy~beam$^{-1}$), 
and IRAS 04381+2540 ($1\sigma$~=~0.100~Jy~beam$^{-1}$, min~=~$-$0.2~Jy~beam$^{-1}$, max~=~1.9~Jy~beam$^{-1}$).}
\label{fig_multiple5}
\end{figure*}

\begin{figure*}
\epsscale{0.95}
\plotone{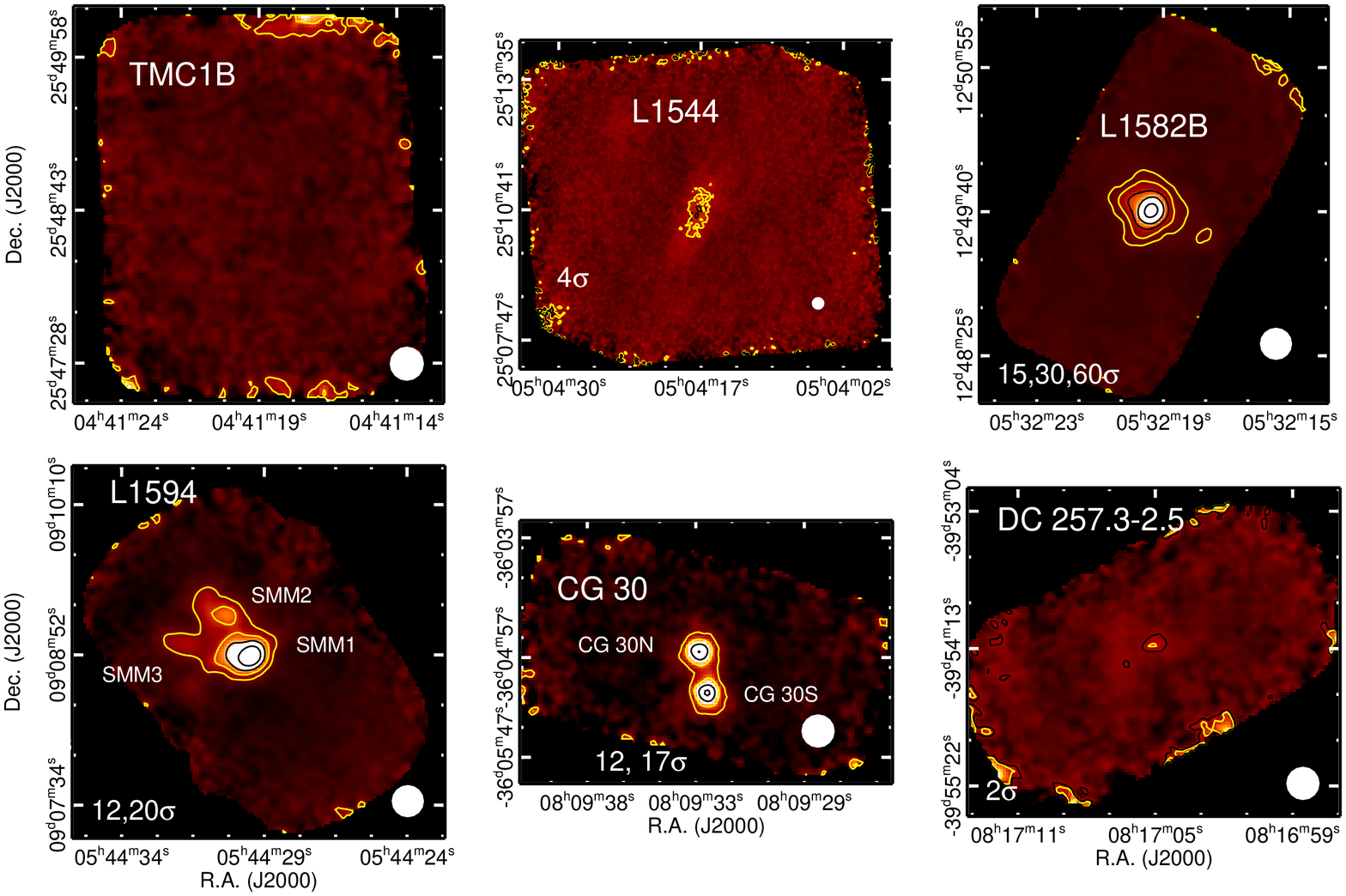}
\caption{Same as Figure \ref{fig_L1448}, except for 
TMC1B ($1\sigma$~=~0.086~Jy~beam$^{-1}$, min~=~$-$0.3~Jy~beam$^{-1}$, max~=~1.1~Jy~beam$^{-1}$), 
L1544 ($1\sigma$~=~0.058~Jy~beam$^{-1}$, min~=~$-$0.2~Jy~beam$^{-1}$, max~=~0.6~Jy~beam$^{-1}$), 
L1582B ($1\sigma$~=~0.123~Jy~beam$^{-1}$, min~=~$-$0.8~Jy~beam$^{-1}$, max~=~3.9~Jy~beam$^{-1}$), 
L1594 ($1\sigma$~=~0.210~Jy~beam$^{-1}$, min~=~$-$0.4~Jy~beam$^{-1}$, max~=~2.8~Jy~beam$^{-1}$), 
CG 30 ($1\sigma$~=~0.452~Jy~beam$^{-1}$, min~=~$-$0.4~Jy~beam$^{-1}$, max~=~4.3~Jy~beam$^{-1}$), 
and DC 257.3-2.5 ($1\sigma$~=~0.067~Jy~beam$^{-1}$, min~=~$-$0.2~Jy~beam$^{-1}$, max~=~0.6~Jy~beam$^{-1}$).}
\label{fig_multiple6}
\end{figure*}

\begin{figure*}
\epsscale{0.95}
\plotone{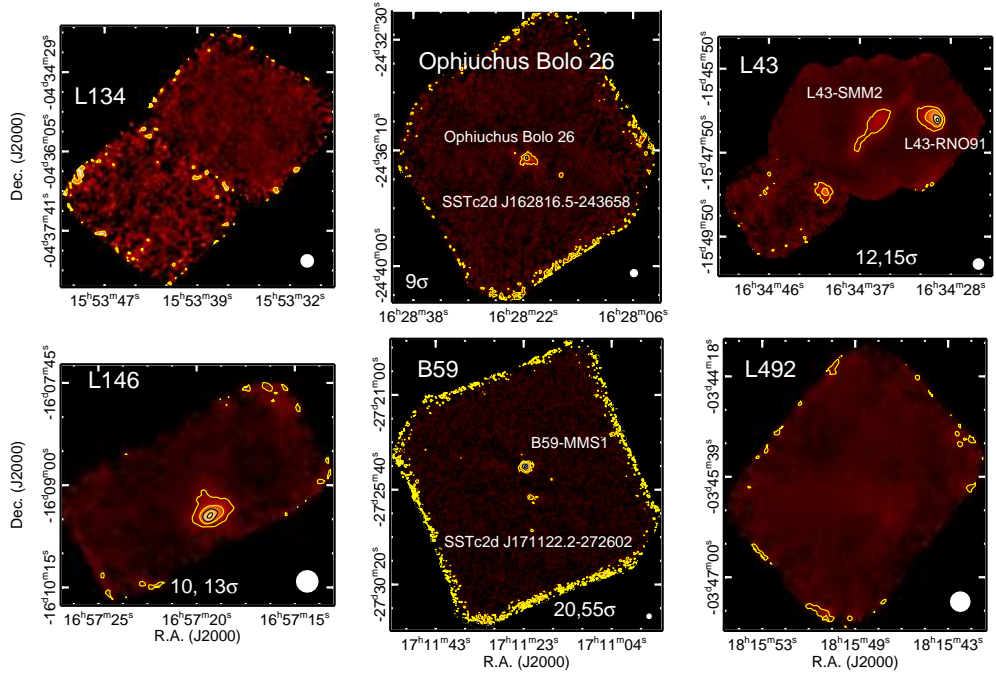}
\caption{Same as Figure \ref{fig_L1448}, except for 
L134 ($1\sigma$~=~0.196~Jy~beam$^{-1}$, min~=~$-$0.3~Jy~beam$^{-1}$, max~=~1.2~Jy~beam$^{-1}$), 
Ophiuchus Bolo 26 ($1\sigma$~=~0.181~Jy~beam$^{-1}$, min~=~$-$0.3~Jy~beam$^{-1}$, max~=~2.2~Jy~beam$^{-1}$), 
L43 ($1\sigma$~=~0.194~Jy~beam$^{-1}$, min~=~$-$0.6~Jy~beam$^{-1}$, max~=~2.9~Jy~beam$^{-1}$), 
L146 ($1\sigma$~=~0.338~Jy~beam$^{-1}$, min~=~$-$0.6~Jy~beam$^{-1}$, max~=~4.7~Jy~beam$^{-1}$), 
B59 ($1\sigma$~=~0.319~Jy~beam$^{-1}$, min~=~$-$0.4~Jy~beam$^{-1}$, max~=~5.0~Jy~beam$^{-1}$), 
and L492 ($1\sigma$~=~0.052~Jy~beam$^{-1}$, min~=~$-$0.2~Jy~beam$^{-1}$, max~=~1.1~Jy~beam$^{-1}$).}
\label{fig_multiple7}
\end{figure*}

\begin{figure*}
\epsscale{0.95}
\plotone{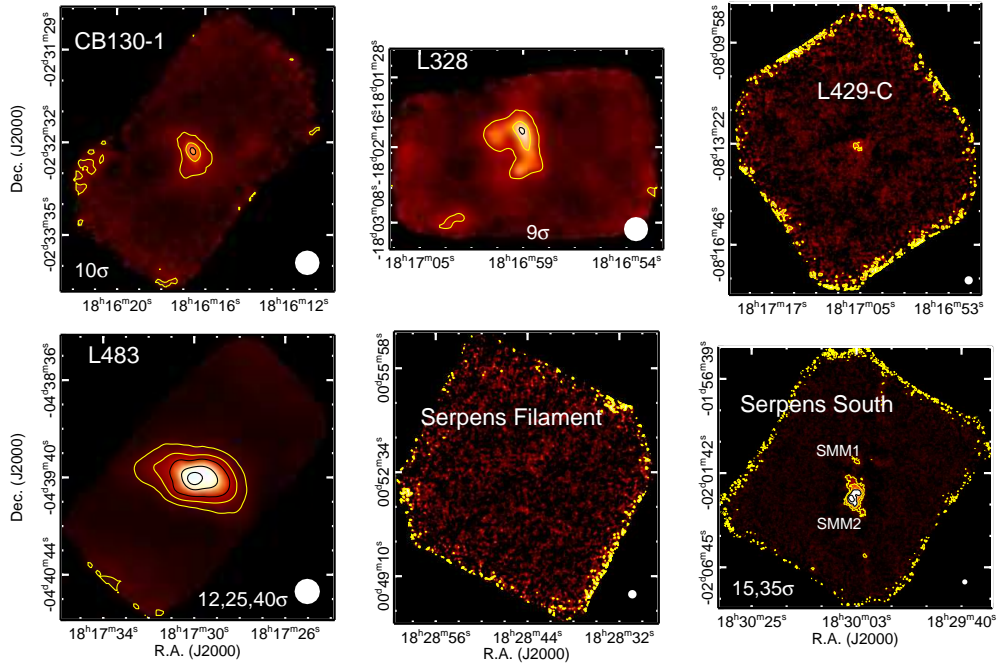}
\caption{Same as Figure \ref{fig_L1448}, except for 
CB130-1 ($1\sigma$~=~0.069~Jy~beam$^{-1}$, min~=~$-$0.1~Jy~beam$^{-1}$, max~=~1.0~Jy~beam$^{-1}$), 
L328 ($1\sigma$~=~0.121~Jy~beam$^{-1}$, min~=~$-$0.3~Jy~beam$^{-1}$, max~=~1.1~Jy~beam$^{-1}$), 
L429-C ($1\sigma$~=~0.117~Jy~beam$^{-1}$, min~=~$-$0.1~Jy~beam$^{-1}$, max~=~0.7~Jy~beam$^{-1}$), 
L483 ($1\sigma$~=~0.251~Jy~beam$^{-1}$, min~=~$-$0.6~Jy~beam$^{-1}$, max~=~6.3~Jy~beam$^{-1}$), 
Serpens Filament ($1\sigma$~=~0.478~Jy~beam$^{-1}$, min~=~$-$0.1~Jy~beam$^{-1}$, max~=~2.0~Jy~beam$^{-1}$), 
and Serpens South ($1\sigma$~=~0.432~Jy~beam$^{-1}$, min~=~$-$0.4~Jy~beam$^{-1}$, max~=~6.5~Jy~beam$^{-1}$).}
\label{fig_multiple8}
\end{figure*}

\begin{figure*}
\epsscale{0.7}
\plotone{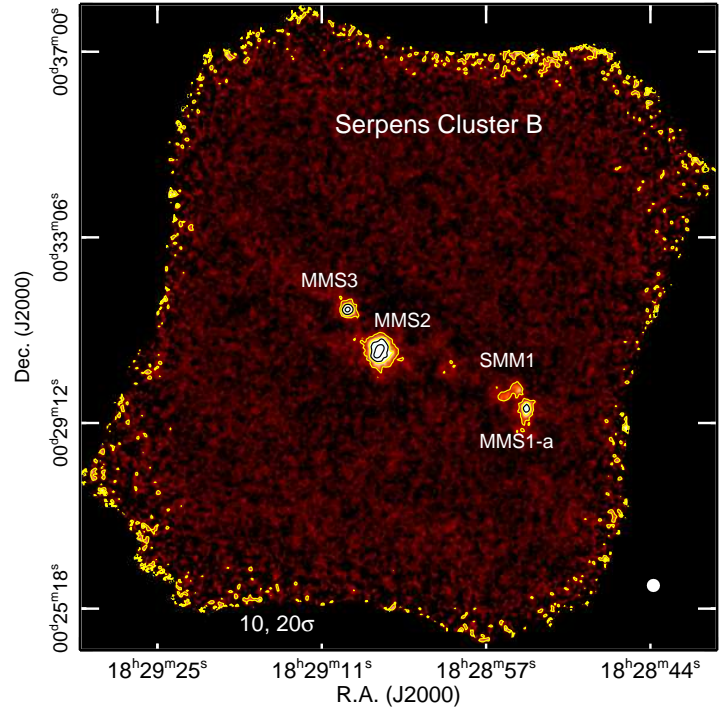}
\caption{Same as Figure \ref{fig_L1448}, except for 
Serpens Cluster B ($1\sigma$~=~0.352~Jy~beam$^{-1}$, min~=~$-$0.4~Jy~beam$^{-1}$, max~=~2.6~Jy~beam$^{-1}$).}
\label{fig_SERPCLUSB}
\end{figure*}

\begin{figure*}
\epsscale{0.7}
\plotone{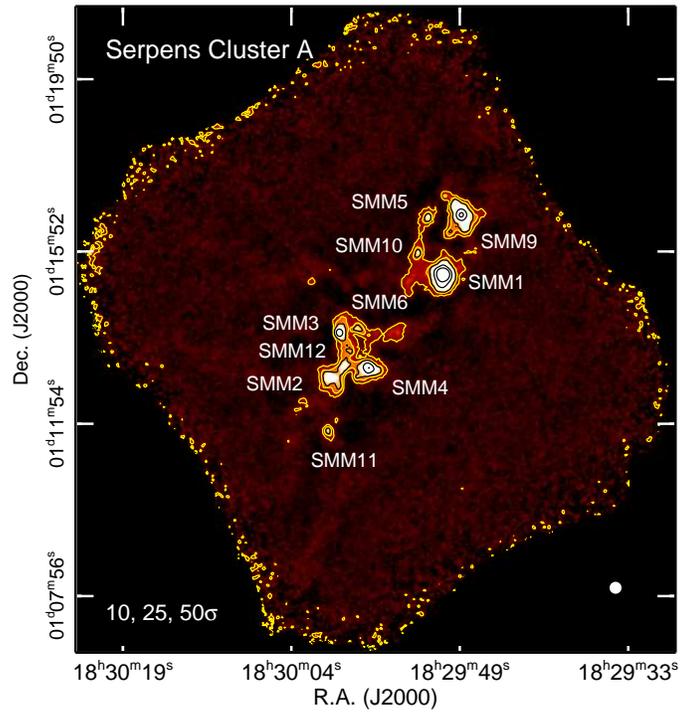}
\caption{Same as Figure \ref{fig_L1448}, except for 
Serpens Cluster A ($1\sigma$~=~0.342~Jy~beam$^{-1}$, min~=~$-$0.7~Jy~beam$^{-1}$, max~=~4.6~Jy~beam$^{-1}$).}
\label{fig_SERPCLUSA}
\end{figure*}

\begin{figure*}
\epsscale{1.0}
\plotone{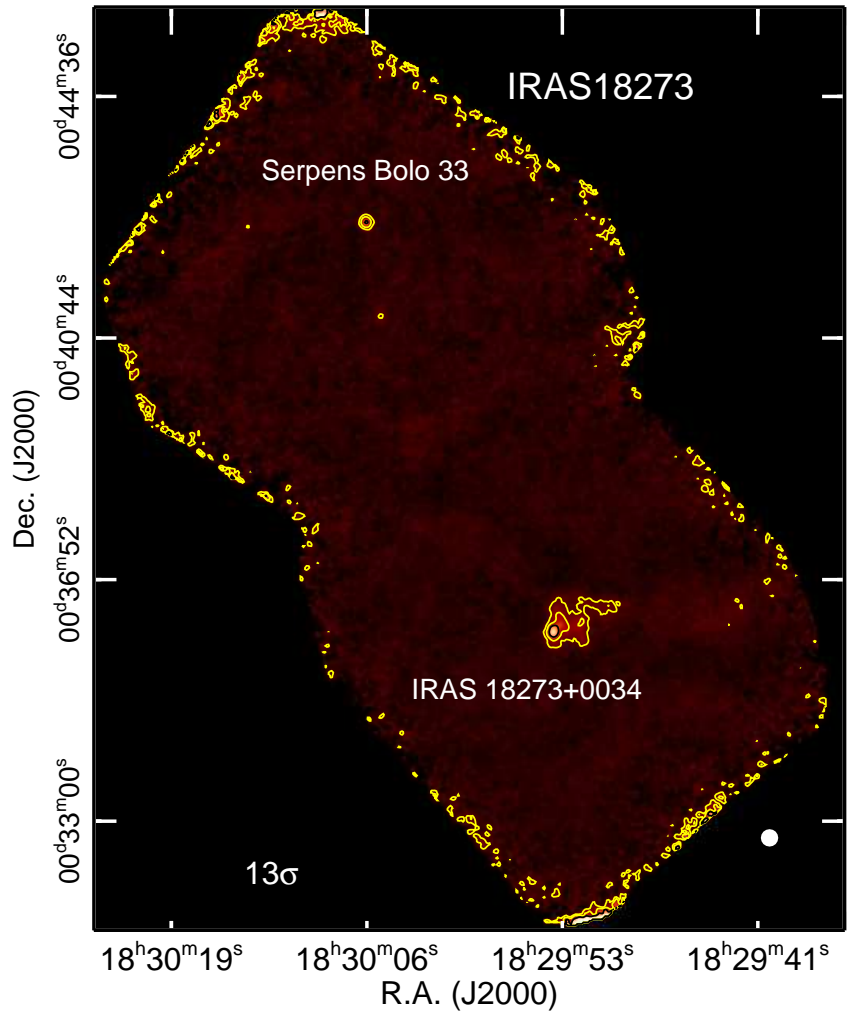}
\caption{Same as Figure \ref{fig_L1448}, except for 
IRAS 18273 ($1\sigma$~=~0.101~Jy~beam$^{-1}$, min~=~$-$0.3~Jy~beam$^{-1}$, max~=~2.2~Jy~beam$^{-1}$).}
\label{fig_IRAS18273}
\end{figure*}

\begin{figure*}
\epsscale{1.0}
\plotone{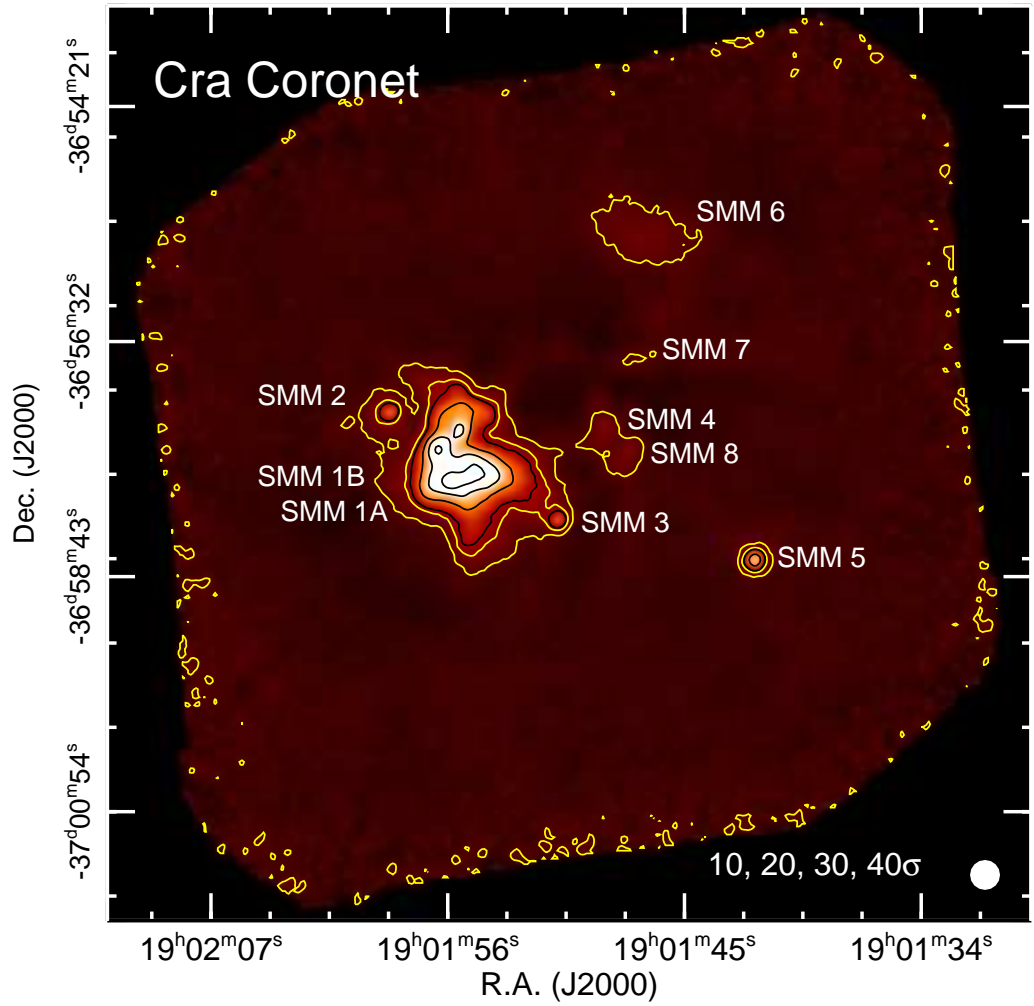}
\caption{Same as Figure \ref{fig_L1448}, except for 
CrA Coronet ($1\sigma$~=~0.445~Jy~beam$^{-1}$, min~=~$-$2.7~Jy~beam$^{-1}$, max~=~13.6~Jy~beam$^{-1}$).}
\label{fig_cracoronet}
\end{figure*}

\begin{figure*}
\epsscale{0.95}
\plotone{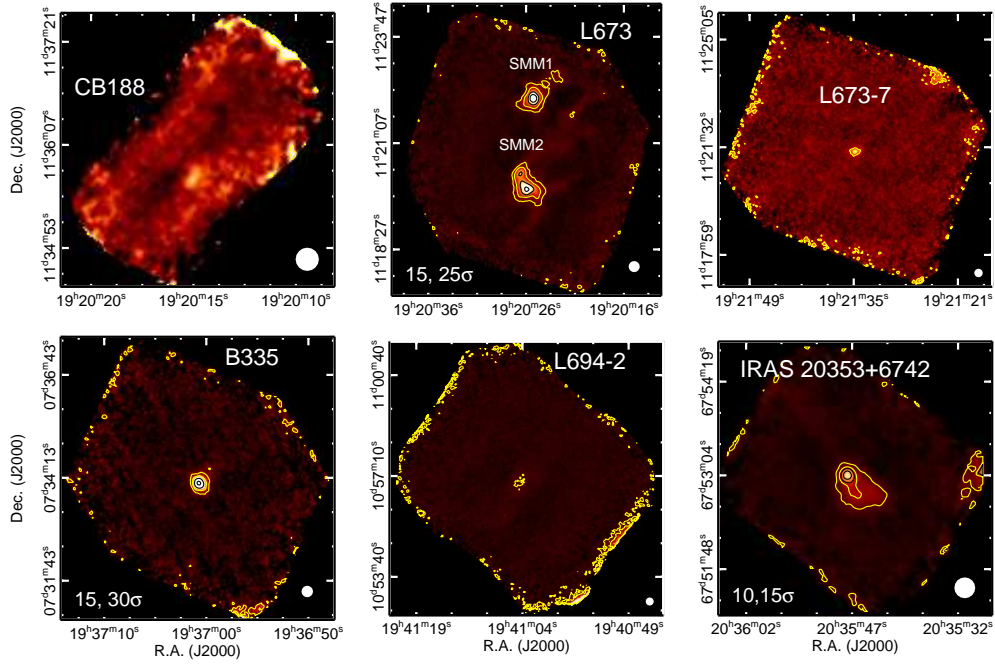}
\caption{Same as Figure \ref{fig_L1448}, except for 
CB188 ($1\sigma$~=~0.261~Jy~beam$^{-1}$, min~=~$-$0.4~Jy~beam$^{-1}$, max~=~0.9~Jy~beam$^{-1}$), 
L673 ($1\sigma$~=~0.114~Jy~beam$^{-1}$, min~=~$-$0.3~Jy~beam$^{-1}$, max~=~2.2~Jy~beam$^{-1}$), 
L673-7 ($1\sigma$~=~0.083~Jy~beam$^{-1}$, min~=~$-$0.3~Jy~beam$^{-1}$, max~=~0.7~Jy~beam$^{-1}$), 
B335 ($1\sigma$~=~0.298~Jy~beam$^{-1}$, min~=~$-$0.4~Jy~beam$^{-1}$, max~=~3.3~Jy~beam$^{-1}$), 
L694-2 ($1\sigma$~=~0.078~Jy~beam$^{-1}$, min~=~$-$0.3~Jy~beam$^{-1}$, max~=~2.2~Jy~beam$^{-1}$), 
and IRAS 20353+6742 ($1\sigma$~=~0.105~Jy~beam$^{-1}$, min~=~$-$0.2~Jy~beam$^{-1}$, max~=~2.0~Jy~beam$^{-1}$).}
\label{fig_multiple9}
\end{figure*}

\begin{figure*}
\epsscale{0.95}
\epsscale{1.0}
\plotone{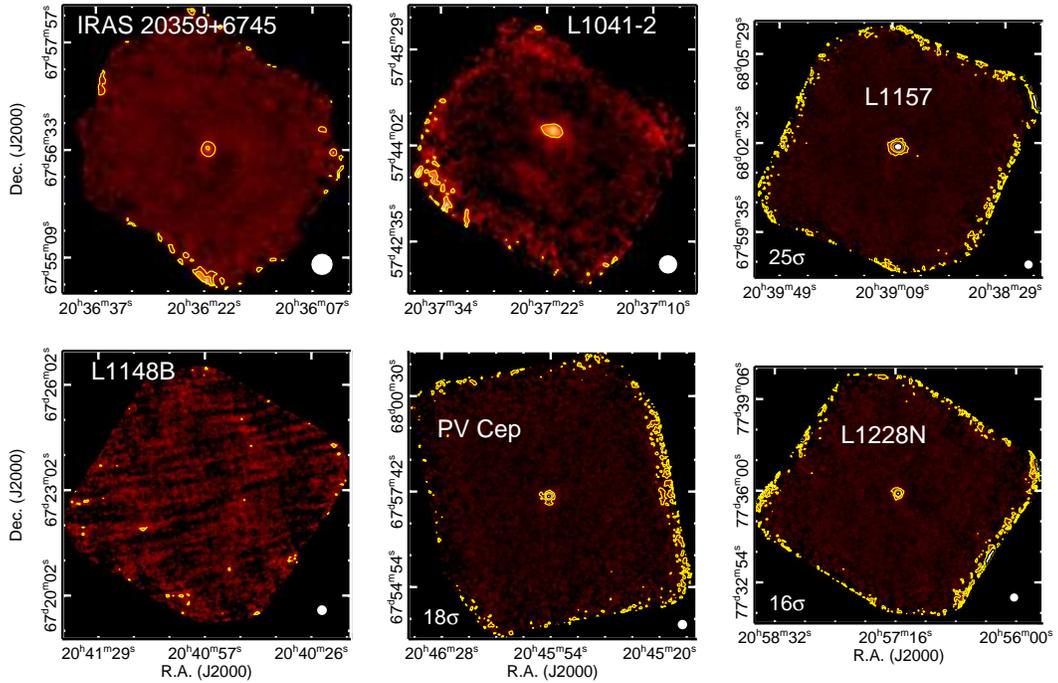}
\caption{Same as Figure \ref{fig_L1448}, except for 
IRAS 20359+6745 ($1\sigma$~=~0.081~Jy~beam$^{-1}$, min~=~$-$0.3~Jy~beam$^{-1}$, max~=~1.1~Jy~beam$^{-1}$), 
L1041-2 ($1\sigma$~=~0.536~Jy~beam$^{-1}$, min~=~$-$0.4~Jy~beam$^{-1}$, max~=~3.2~Jy~beam$^{-1}$), 
L1157 ($1\sigma$~=~0.161~Jy~beam$^{-1}$, min~=~$-$0.4~Jy~beam$^{-1}$, max~=~3.6~Jy~beam$^{-1}$), 
L1148B ($1\sigma$~=~0.127~Jy~beam$^{-1}$, min~=~$-$0.1~Jy~beam$^{-1}$, max~=~0.8~Jy~beam$^{-1}$), 
PV Cep ($1\sigma$~=~0.242~Jy~beam$^{-1}$, min~=~$-$0.4~Jy~beam$^{-1}$, max~=~3.5~Jy~beam$^{-1}$), 
and L1228N ($1\sigma$~=~0.231~Jy~beam$^{-1}$, min~=~$-$0.4~Jy~beam$^{-1}$, max~=~3.6~Jy~beam$^{-1}$).}
\label{fig_multiple10}
\end{figure*}

\begin{figure*}
\epsscale{0.95}
\epsscale{1.0}
\plotone{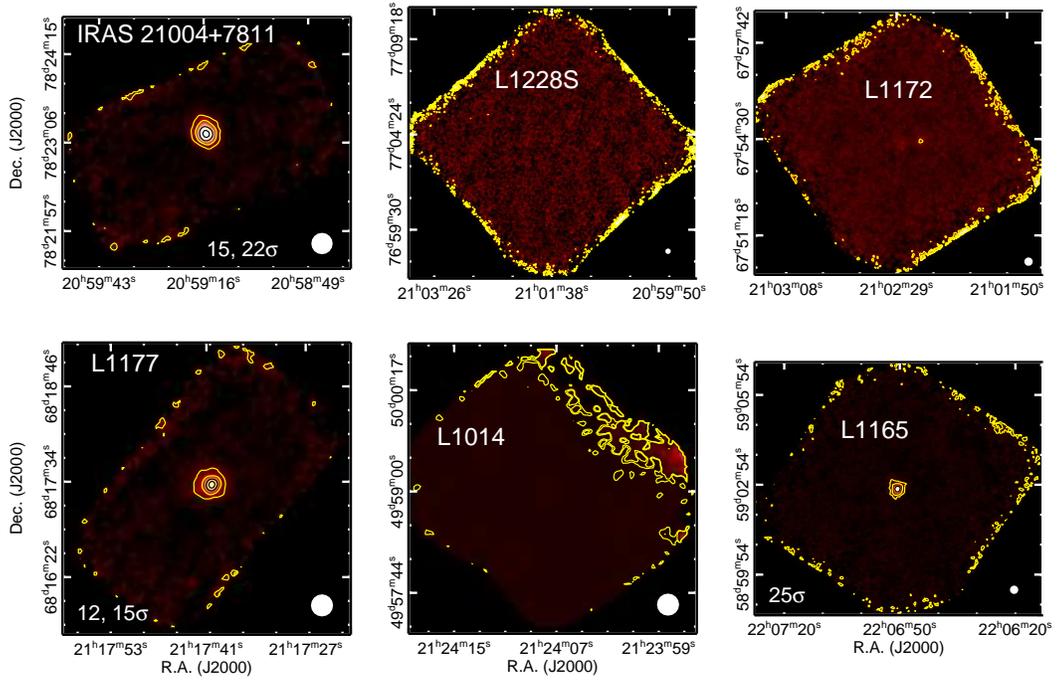}
\caption{Same as Figure \ref{fig_L1448}, except for 
IRAS 21004+7811 ($1\sigma$~=~0.466~Jy~beam$^{-1}$, min~=~$-$0.4~Jy~beam$^{-1}$, max~=~6.9~Jy~beam$^{-1}$), 
L1228S ($1\sigma$~=~0.313~Jy~beam$^{-1}$, min~=~$-$0.4~Jy~beam$^{-1}$, max~=~2.3~Jy~beam$^{-1}$), 
L1172 ($1\sigma$~=~0.154~Jy~beam$^{-1}$, min~=~$-$0.4~Jy~beam$^{-1}$, max~=~2.2~Jy~beam$^{-1}$), 
L1177 ($1\sigma$~=~0.405~Jy~beam$^{-1}$, min~=~$-$0.4~Jy~beam$^{-1}$, max~=~6.3~Jy~beam$^{-1}$), 
L1014 ($1\sigma$~=~0.008~Jy~beam$^{-1}$, min~=~$-$0.2~Jy~beam$^{-1}$, max~=~1.6~Jy~beam$^{-1}$), 
and L1165 ($1\sigma$~=~0.163~Jy~beam$^{-1}$, min~=~$-$0.3~Jy~beam$^{-1}$, max~=~4.7~Jy~beam$^{-1}$).}
\label{fig_multiple11}
\end{figure*}

\begin{figure*}
\epsscale{0.95}
\epsscale{1.0}
\plotone{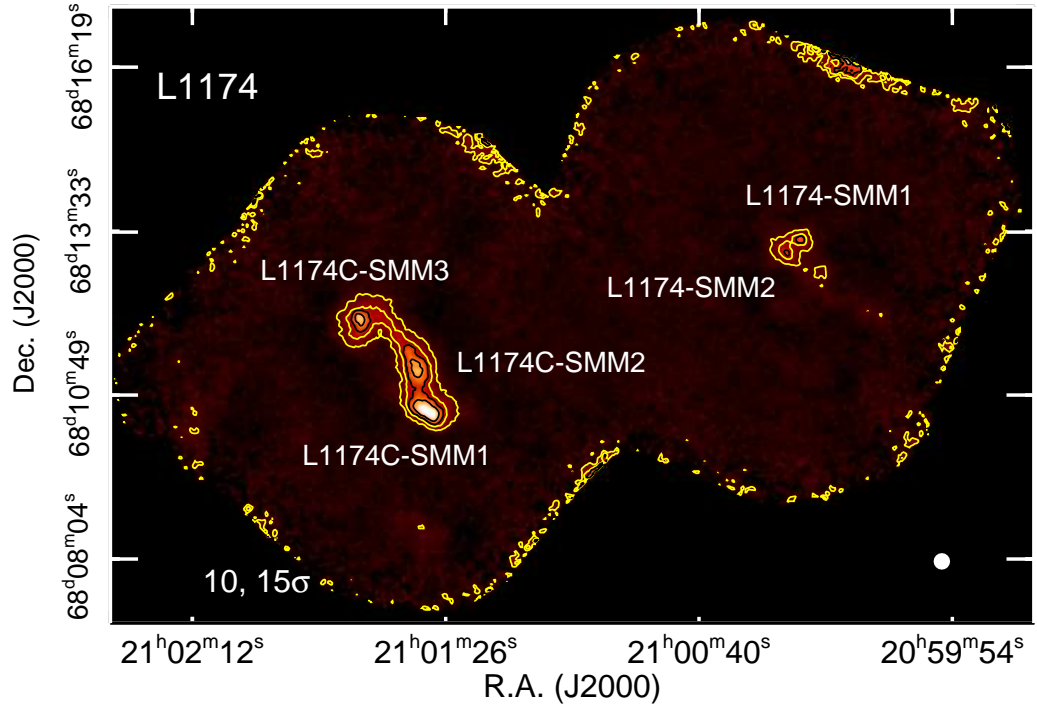}
\caption{Same as Figure \ref{fig_L1448}, except for 
L1174 ($1\sigma$~=~0.141~Jy~beam$^{-1}$, min~=~$-$0.3~Jy~beam$^{-1}$, max~=~2.9~Jy~beam$^{-1}$).}
\label{fig_L1174}
\end{figure*}

\begin{figure*}[hbt!]
\epsscale{0.95}
\epsscale{1.0}
\plotone{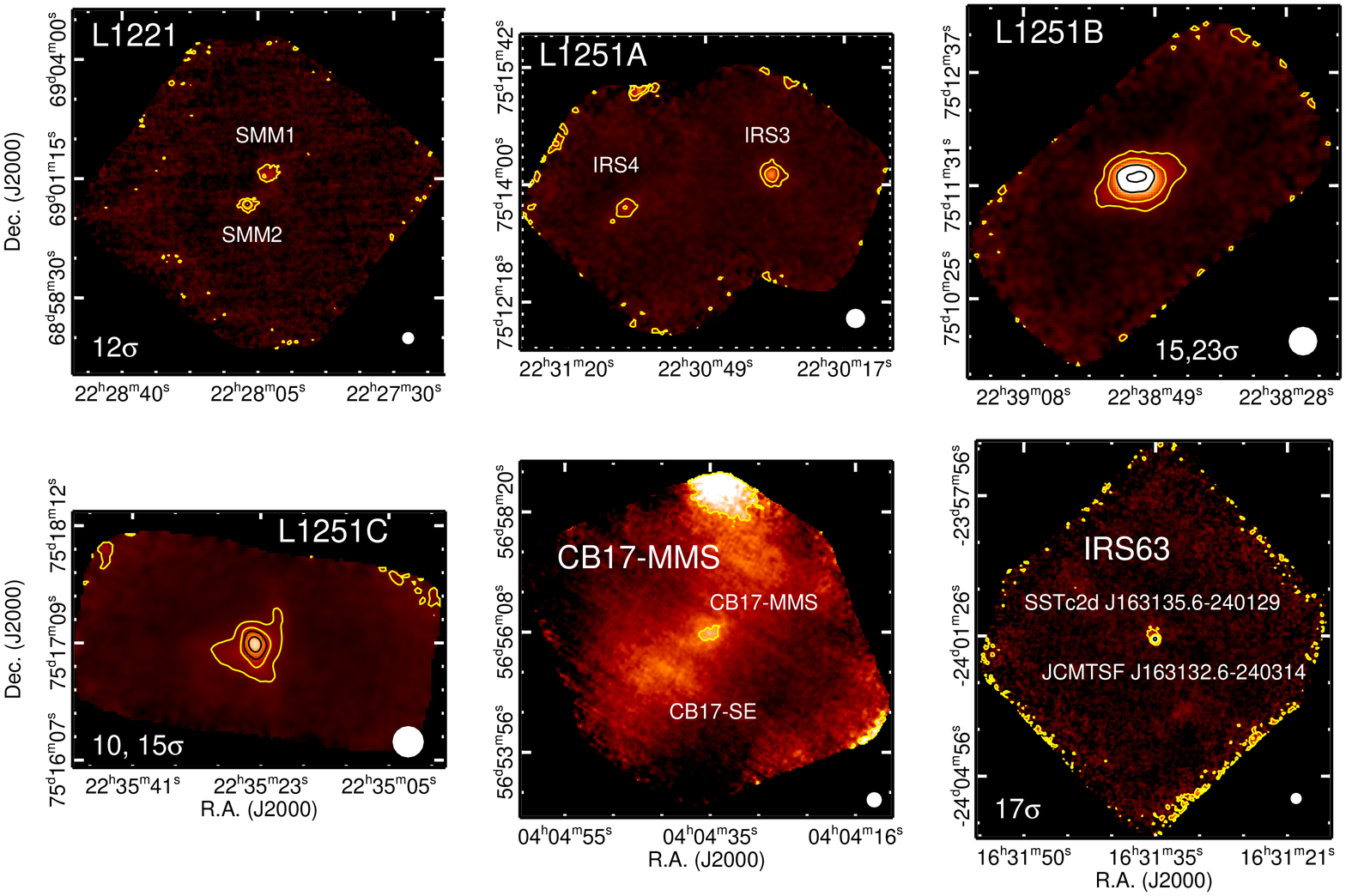}
\caption{Same as Figure \ref{fig_L1448}, except for 
L1221 ($1\sigma$~=~0.250~Jy~beam$^{-1}$, min~=~$-$0.4~Jy~beam$^{-1}$, max~=~3.7~Jy~beam$^{-1}$), 
L1251A ($1\sigma$~=~0.144~Jy~beam$^{-1}$, min~=~$-$0.4~Jy~beam$^{-1}$, max~=~2.8~Jy~beam$^{-1}$), 
L1251B ($1\sigma$~=~0.278~Jy~beam$^{-1}$, min~=~$-$0.5~Jy~beam$^{-1}$, max~=~4.4~Jy~beam$^{-1}$), 
L1251C ($1\sigma$~=~0.240~Jy~beam$^{-1}$, min~=~$-$0.8~Jy~beam$^{-1}$, max~=~4.9~Jy~beam$^{-1}$), 
CB17-MMS ($1\sigma$~=~0.405~Jy~beam$^{-1}$, min~=~$-$0.1~Jy~beam$^{-1}$, max~=~1.7~Jy~beam$^{-1}$), 
and IRS63 ($1\sigma$~=~0.220~Jy~beam$^{-1}$, min~=~$-$0.2~Jy~beam$^{-1}$, max~=~2.0~Jy~beam$^{-1}$).}
\label{fig_multiple12}
\end{figure*}

Figures \ref{fig_L1448} -- \ref{fig_multiple12} show contour maps overlaid on 
images for each of the 81 regions listed in Table 
\ref{tab_properties1}. Since the source extraction routine uses a local noise 
measurement but the contours are plotted using a global noise measurement, 
weak sources in low-noise regions may lack associated 3$\sigma$ 
contours.  The images are of the maps reduced without the extended 
emission flag. Most maps are displayed in six-panel figures 
except for the largest and most crowded maps, which are instead displayed as 
one-panel figures.  Maps with multiple sources have each source labeled.  
All of the reduced FITS files are available in calibrated units of 
Jy~beam$^{-1}$ following the calibration procedures described above; versions 
with and without the extended flag can be accessed through the Data Behind the 
Figures (DBF) feature of the journal.

We detect a total of 164 sources in the 81 maps listed in Table 
\ref{tab_properties1} and presented in Figures \ref{fig_L1448} -- 
\ref{fig_multiple12}.  
Table \ref{tab_sources} lists, for each detected source, 
the name of the source, the map in which the source is covered, the peak 
position of the source, the deconvolved angular source radius as determined by Bolocat 
\citep[see][for details]{rosolowsky2010:bolocat}, the peak intensity, 
the flux density in a 20$''$ diameter aperture, the flux density in a 
40$''$ diameter aperture, and a flag noting whether each source is starless 
or protostellar, determined mostly from a search for an infrared point source 
in {\it Spitzer Space Telescope} images from the c2d 
\citep[e.g.,][]{evans2009:c2d} and Gould Belt \citep[e.g.,][]{dunham2015:gb} 
legacy projects.  The peak intensities are given in units of Jy beam$^{-1}$ 
(for the SHARC-II beam, 1 Jy beam$^{-1}$ = 519.7 MJy sr$^{-1}$).

\clearpage

\begin{figure*}
\epsscale{1.0}
\plotone{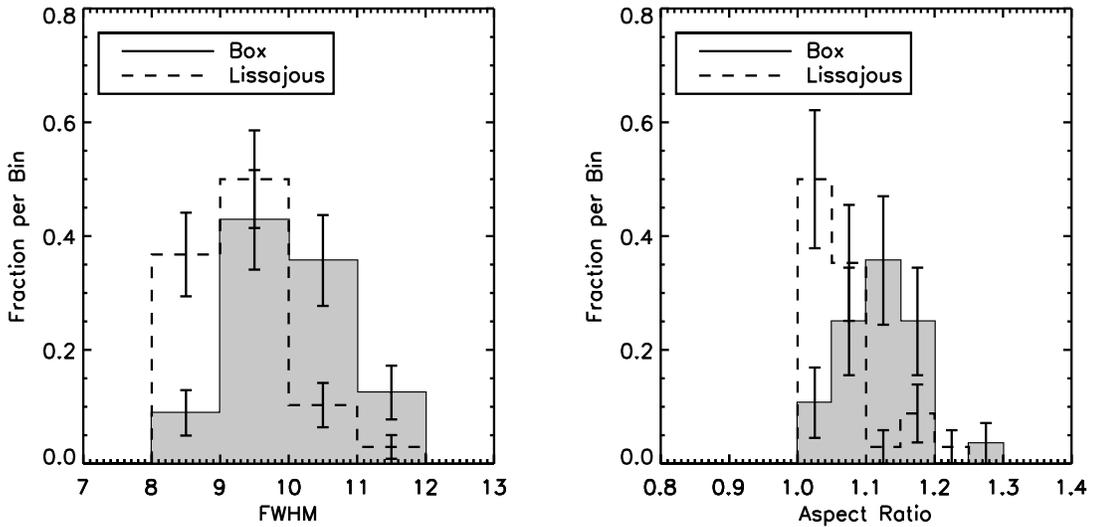}
\caption{\label{fig_boxlisscal}Comparison of the FWHM (left) and aspect ratios 
(right) of the beams derived from the Lissajous (dashed histograms) and box 
(filled histograms) calibration scans.}
\end{figure*}

\clearpage

\section{Effects of Scan Mode}\label{sec_mode}

As noted in Section \ref{sec_observations}, all observations prior to 2006 
were taken in the Lissajous scan mode. However, over the course of taking 
observations we were made aware of a different observing mode, the box scan 
mode, that might yield better results.  In particular, maps observed in the 
box scan mode have much larger areas than those observed in the Lissajous 
mode.  Since all emission on scales larger than the map size will be treated 
as sky emission and removed by CRUSH, the larger map areas provided by the box 
scans may provide better recovery of extended emission.  The use of the box 
scan mode also allows us to map larger areas in reasonable amounts of time.  
Thus, all science data obtained during and after the July 2008 observing run 
were obtained exclusively in the box scan mode, with observations between 
December 2006 and July 2008 obtained in both modes for testing purposes.

Calibration scans were taken with the Lissajous mode, even after 
July 2008, since one Lissajous mode scan can be completed in much less time 
than one box scan.  To justify this choice, we obtained several calibration 
scans in the box-scan observing mode.  The calibration factors obtained from 
these scans are listed in Table \ref{tab_calibrators3}, using the same 
method as above for the Lissajous calibration scans.  Averaged over all of the 
box calibration scans, we calculate C$_{\rm beam} = 1.63 \pm 0.15$, 
C$_{\rm 20} = 0.034 \pm 0.004$, and C$_{\rm 40} = 0.025 \pm 0.003$, where the 
uncertainties are the standard deviations.  Comparison to Table 
\ref{tab_calibrators2} shows that the mean Lissajous and box-scan calibration 
factors agree to within 10\%, well within the overall calibration uncertainty 
of 25\%, indicating that any dependence on observing mode in the SHARC-II 
instrument calibration has a negligible impact on our results.

We also compared the SHARC-II beams resulting from the Lissajous and box scans. 
Figure \ref{fig_boxlisscal} compares the size and shape of the beams derived 
from the Lissajous and box calibration scans.  The beam profiles are derived 
by deconvolving the measured properties (sizes and elongations) of the 
calibration sources with their known, intrinsic properties.  
For the beam FWHM, the means 
and standard deviations of the means are 9.3$''$ $\pm$ 0.2$''$ for the 
Lissajous scans and 10.0$''$ $\pm$ 0.2$''$ for the box scans.  For the beam 
aspect ratio, the means and standard deviations of the means are 1.06 $\pm$ 
0.01 for the Lissajous scans and 1.12 $\pm$ 0.01 for the box scans.  Thus, 
observations in the box scan mode have a beam profile that is, on average, 
8\% larger and 
6\% more elongated than observations in the Lissajous mode.  While the
degradation of the beam profile in the box scan mode is a statistically robust 
finding, resulting from both the faster scan rates used by the box scan 
modes and astrometric errors in the SHARC-II pixel plate scale that build 
up over larger scan areas, the overall effects are less than 10\% and have 
no significant impact on our results.

\begin{figure*}
\epsscale{0.95}
\plotone{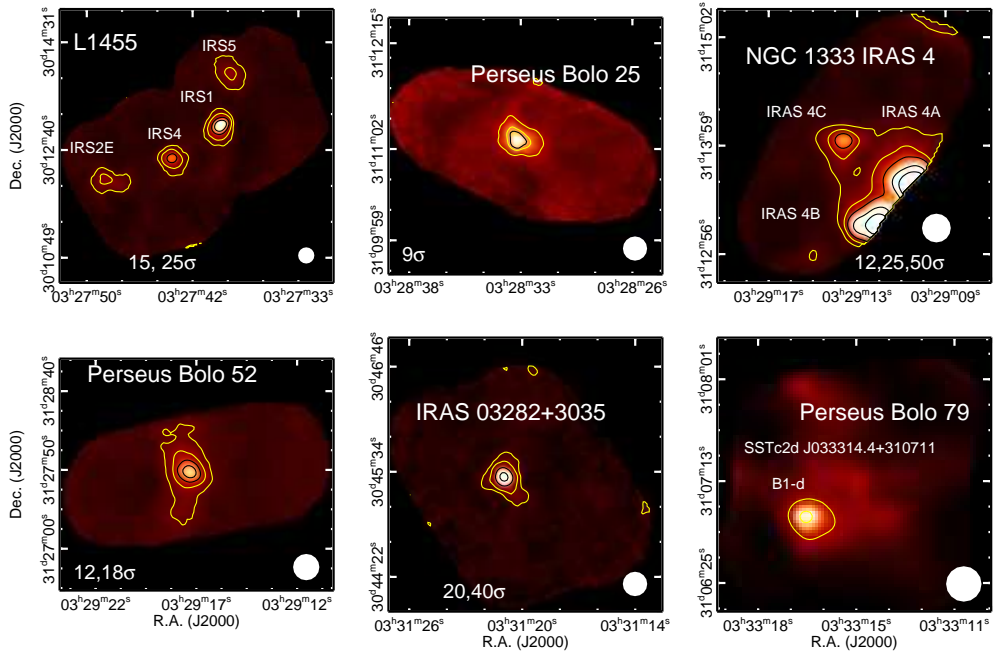}
\caption{SHARC-II 350 $\mu$m Lissajous maps of the targets listed in Table 
\ref{tab_lissajous} (targets observed in both the Lissajous and box-scan 
observing modes), here 
L1455 ($1\sigma$~=~0.137~Jy~beam$^{-1}$, min~=~$-$0.6~Jy~beam$^{-1}$, max~=~4.0~Jy~beam$^{-1}$), 
Perseus Bolo 25 ($1\sigma$~=~0.040~Jy~beam$^{-1}$, min~=~$-$0.2~Jy~beam$^{-1}$, max~=~0.6~Jy~beam$^{-1}$), 
NGC 1333 IRAS 4 ($1\sigma$~=~0.341~Jy~beam$^{-1}$, min~=~$-$0.6~Jy~beam$^{-1}$, max~=~4.8~Jy~beam$^{-1}$), 
Perseus Bolo 52 ($1\sigma$~=~0.098~Jy~beam$^{-1}$, min~=~$-$0.5~Jy~beam$^{-1}$, max~=~2.3~Jy~beam$^{-1}$), 
IRAS 03282+3035 ($1\sigma$~=~0.169~Jy~beam$^{-1}$, min~=~$-$0.6~Jy~beam$^{-1}$, max~=~5.0~Jy~beam$^{-1}$), 
and Perseus Bolo 79 ($1\sigma$~=~0.358~Jy~beam$^{-1}$, min~=~$-$0.2~Jy~beam$^{-1}$, max~=~3.1~Jy~beam$^{-1}$).  
Maps with multiple sources have each source labeled. The beam size is shown at 
the lower right of each map.  The two yellow contour levels are plotted at 
3$\sigma$ and 7$\sigma$.  Additional contours are plotted in black, 
with the levels chosen manually for optimal visual display.  These levels are 
printed in white text at the bottom of each panel, with no text indicating 
no additional black contours are plotted.  Emission seen toward the edges of 
the maps is not reliable and should be ignored.  The color scaling uses a 
linear intensity scale; see Figure \ref{fig_scalebar} for a normalized version 
of the adopted color scale bar.}
\label{fig_lissmultiple1}
\end{figure*}

\begin{figure*}
\epsscale{0.95}
\plotone{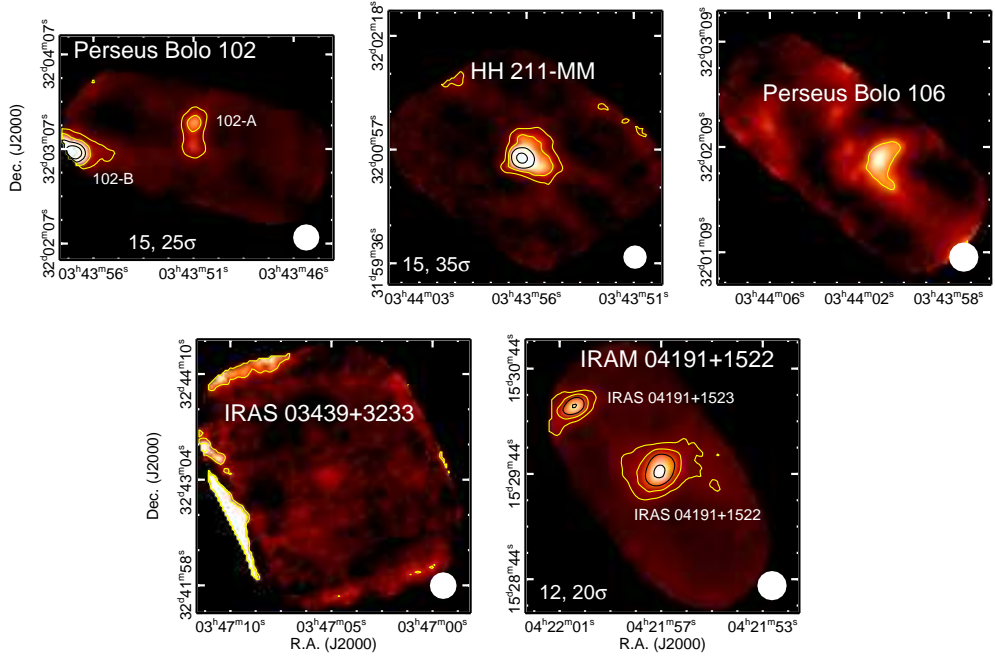}
\caption{Same as Figure \ref{fig_lissmultiple1} (Lissajous observations of 
targets observed in both models), except for 
Perseus Bolo 102 ($1\sigma$~=~0.195~Jy~beam$^{-1}$, min~=~$-$0.3~Jy~beam$^{-1}$, max~=~2.5~Jy~beam$^{-1}$), 
HH 211-MM ($1\sigma$~=~0.226~Jy~beam$^{-1}$, min~=~$-$0.3~Jy~beam$^{-1}$, max~=~2.8~Jy~beam$^{-1}$), 
Perseus Bolo 106 ($1\sigma$~=~0.287~Jy~beam$^{-1}$, min~=~$-$0.3~Jy~beam$^{-1}$, max~=~1.3~Jy~beam$^{-1}$), 
IRAS 03439+3233 ($1\sigma$~=~0.152~Jy~beam$^{-1}$, min~=~$-$0.2~Jy~beam$^{-1}$, max~=~1.0~Jy~beam$^{-1}$), 
and IRAM 04191+1522 ($1\sigma$~=~0.119~Jy~beam$^{-1}$, min~=~$-$0.4~Jy~beam$^{-1}$, max~=~1.7~Jy~beam$^{-1}$).}
\label{fig_lissmultiple2}
\end{figure*}

\begin{figure*}
\epsscale{0.95}
\plotone{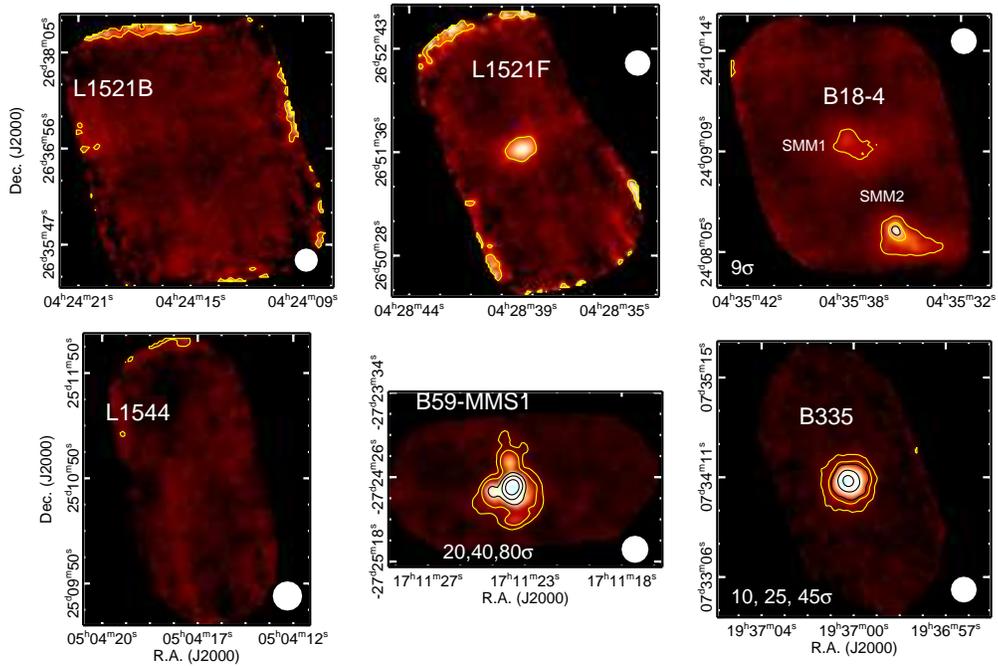}
\caption{Same as Figure \ref{fig_lissmultiple1} (Lissajous observations of 
targets observed in both models), except for 
L1521B ($1\sigma$~=~0.135~Jy~beam$^{-1}$, min~=~$-$0.3~Jy~beam$^{-1}$, max~=~1.1~Jy~beam$^{-1}$), 
L1521F ($1\sigma$~=~0.187~Jy~beam$^{-1}$, min~=~$-$0.3~Jy~beam$^{-1}$, max~=~1.1~Jy~beam$^{-1}$), 
B18-4 ($1\sigma$~=~0.120~Jy~beam$^{-1}$, min~=~$-$0.3~Jy~beam$^{-1}$, max~=~1.2~Jy~beam$^{-1}$), 
L1544 ($1\sigma$~=~0.071~Jy~beam$^{-1}$, min~=~$-$0.1~Jy~beam$^{-1}$, max~=~0.9~Jy~beam$^{-1}$), 
B59-MMS1 ($1\sigma$~=~0.200~Jy~beam$^{-1}$, min~=~$-$0.6~Jy~beam$^{-1}$, max~=~4.2~Jy~beam$^{-1}$), 
and B335 ($1\sigma$~=~0.223~Jy~beam$^{-1}$, min~=~$-$0.5~Jy~beam$^{-1}$, max~=~4.4~Jy~beam$^{-1}$).}
\label{fig_lissmultiple3}
\end{figure*}

\begin{figure*}
\epsscale{0.95}
\plotone{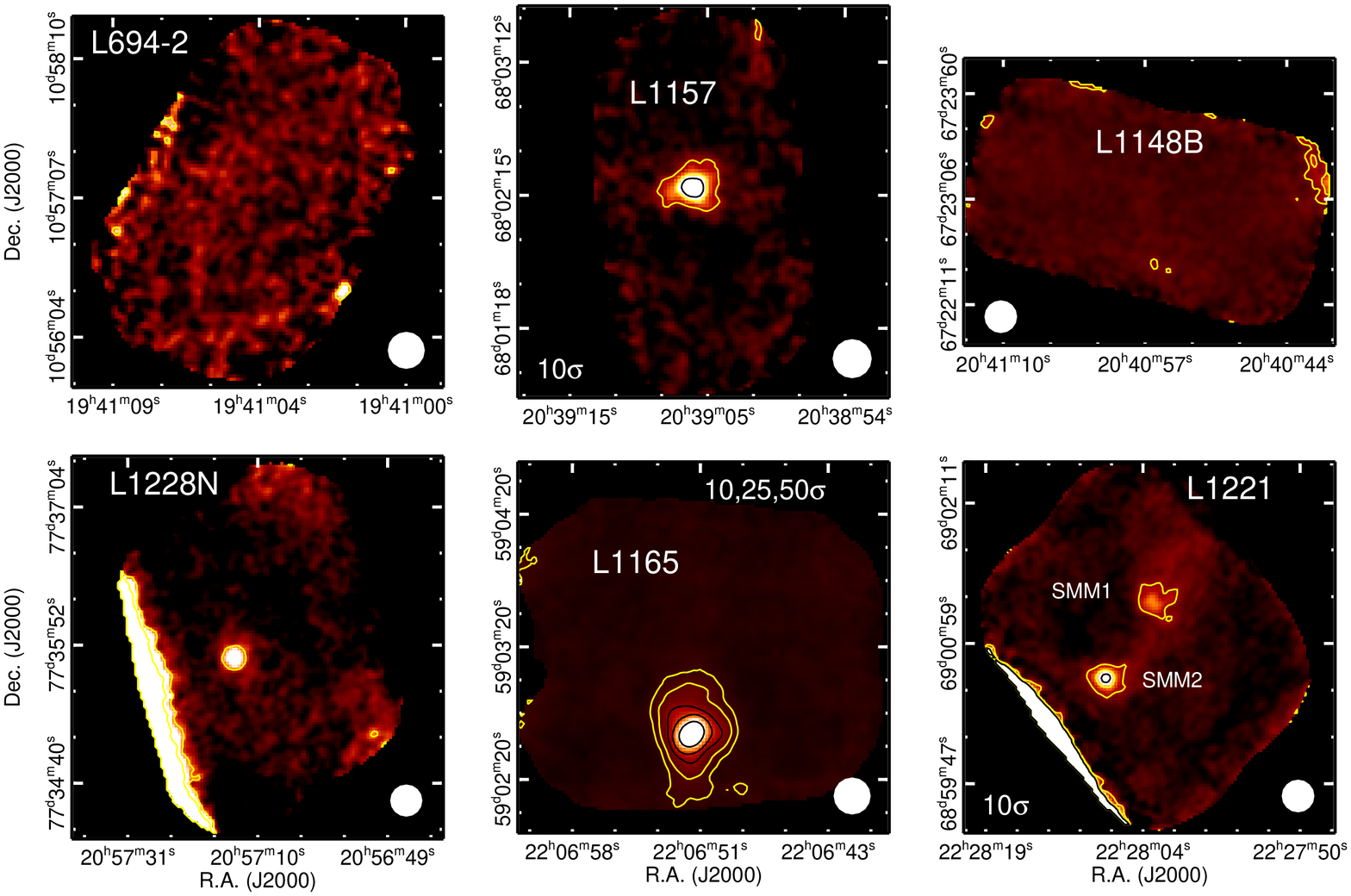}
\caption{Same as Figure \ref{fig_lissmultiple1} (Lissajous observations of 
targets observed in both models), except for 
L694-2 ($1\sigma$~=~0.257~Jy~beam$^{-1}$, min~=~$-$0.3~Jy~beam$^{-1}$, max~=~1.1~Jy~beam$^{-1}$), 
L1157 ($1\sigma$~=~0.631~Jy~beam$^{-1}$, min~=~$-$0.3~Jy~beam$^{-1}$, max~=~5.0~Jy~beam$^{-1}$), 
L1148B ($1\sigma$~=~0.063~Jy~beam$^{-1}$, min~=~$-$0.3~Jy~beam$^{-1}$, max~=~1.2~Jy~beam$^{-1}$), 
L1228N ($1\sigma$~=~1.023~Jy~beam$^{-1}$, min~=~$-$0.3~Jy~beam$^{-1}$, max~=~4.8~Jy~beam$^{-1}$), 
L1165 ($1\sigma$~=~0.087~Jy~beam$^{-1}$, min~=~$-$0.4~Jy~beam$^{-1}$, max~=~3.9~Jy~beam$^{-1}$), 
and L1221 ($1\sigma$~=~0.214~Jy~beam$^{-1}$, min~=~$-$0.2~Jy~beam$^{-1}$, max~=~1.9~Jy~beam$^{-1}$).}
\label{fig_lissmultiple4}
\end{figure*}

\begin{figure*}
\epsscale{1.2}
\plotone{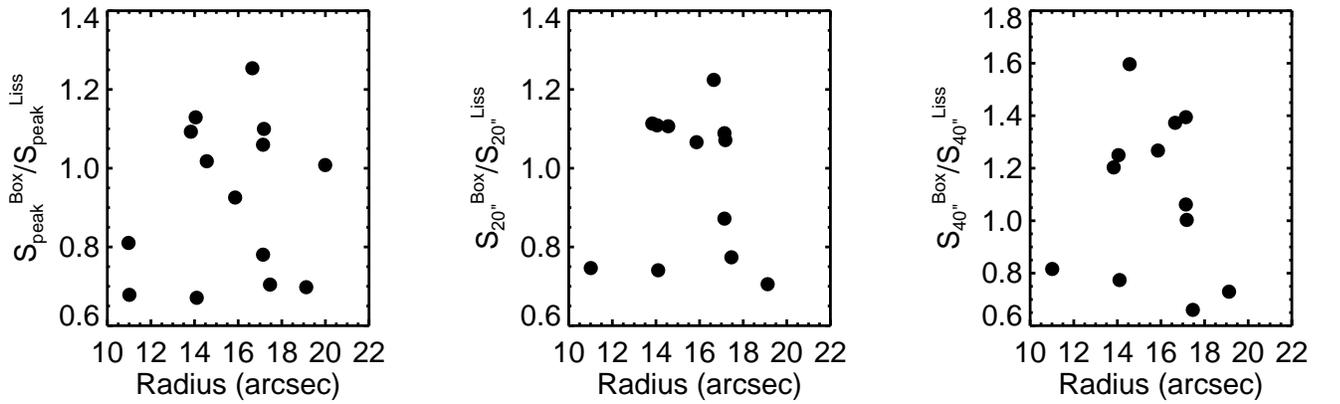}
\caption{\label{fig_radboxliss}Ratios of the flux densities calculated in maps 
observed in the box mode to those observed in the Lissajous mode, plotted 
versus the radius of each source determined from the box scan observations.  
The ratios of the peak intensities, flux densities in 20$''$ diameter 
apertures, and flux densities in 40$''$ diameter apertures are plotted in the 
left, middle, and right panels, respectively.}
\end{figure*}

\begin{figure*}
\epsscale{1.2}
\plotone{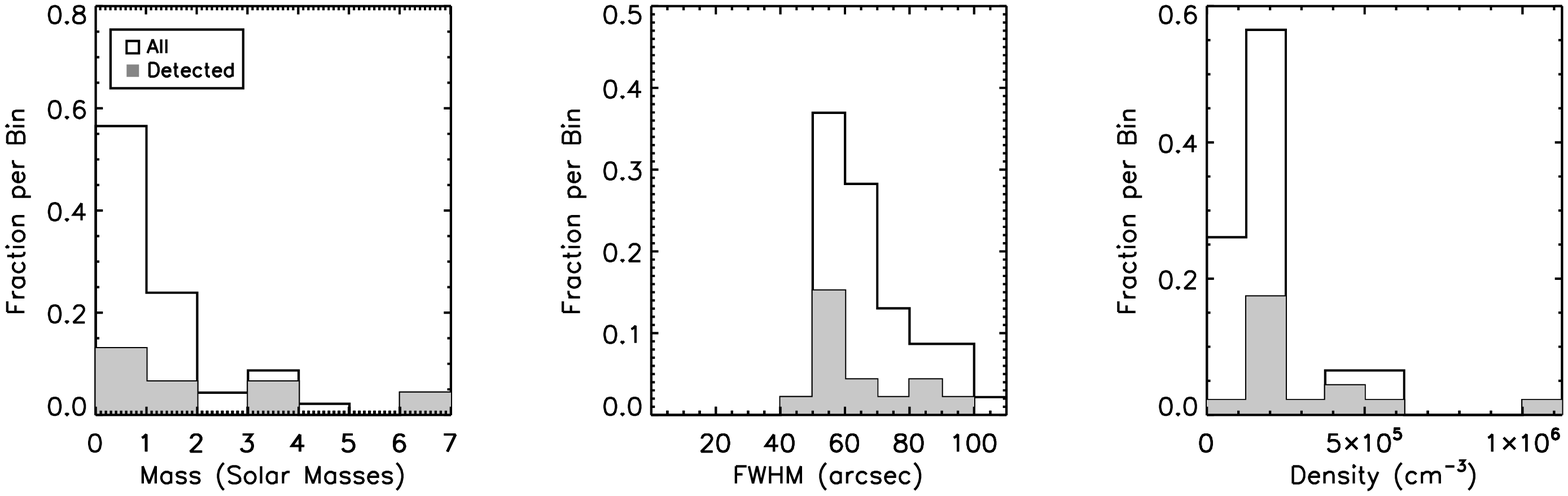}
\caption{\label{fig_starless}Histograms showing the masses, FWHM, and mean 
densities of the starless cores identified in 1.1 mm Bolocam surveys of 
Perseus \citep{1MMSurveyPers}, Ophiuchus \citep{1MMSurveyOph}, and Serpens 
\citep{1MMSurveySerp} and covered in our SHARC-II maps.  The solid, unfilled 
histogram shows the properties of all (detected $+$ undetected) starless cores 
while the shaded histogram shows the properties of the starless cores 
detected in our SHARC-II survey.}
\end{figure*}

\begin{figure}
\epsscale{1.2}
\plotone{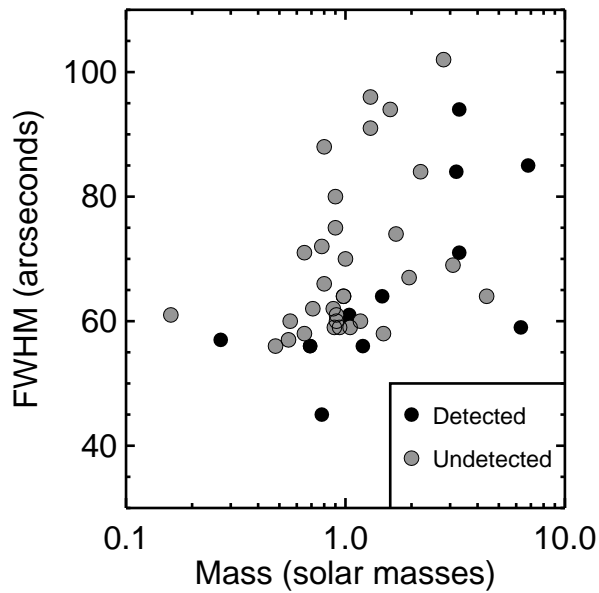}
\caption{\label{fig_starless2}FWHM size plotted versus mass for the starless 
cores identified in 1.1 mm Bolocam surveys of 
Perseus \citep{1MMSurveyPers}, Ophiuchus \citep{1MMSurveyOph}, and Serpens 
\citep{1MMSurveySerp} and covered in our SHARC-II maps.  The black points 
show those starless cores detected in our maps, whereas the light gray 
points show those cores that are not detected in our maps.}
\end{figure}

\begin{figure}
\epsscale{1.2}
\plotone{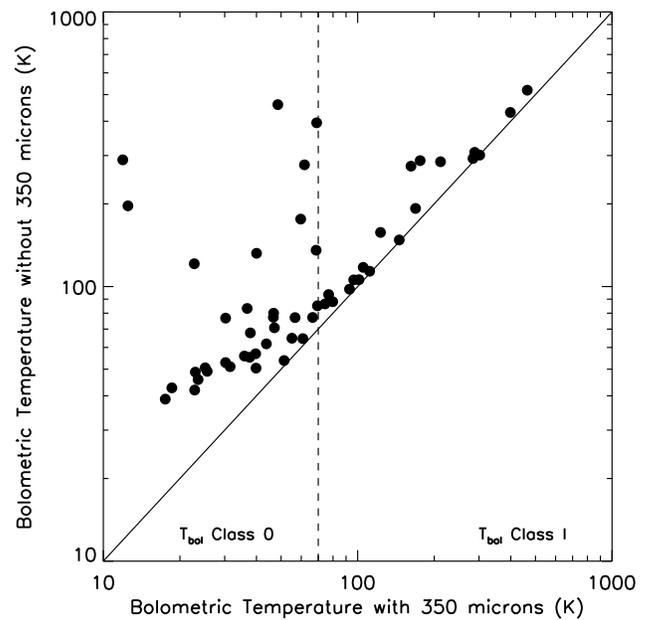}
\caption{\label{fig_classification_tbol}\tbol\ measured without 350 \um\ 
SHARC-II observations included in the SEDs plotted versus \tbol\ measured with 
350 \um\ SHARC-II observations included.  The other data included in the SEDs 
used to calculate both values of \tbol\ are at wavelengths between 
$3.6-70$~\um, and at 1.1~mm (see text for details).  The solid line shows the 
one-to-one line, and the dashed line marks the Class 0/I boundary in \tbol\ as 
defined by \citet{chen1995:tbol}.}
\end{figure}

\begin{figure}
\epsscale{1.2}
\plotone{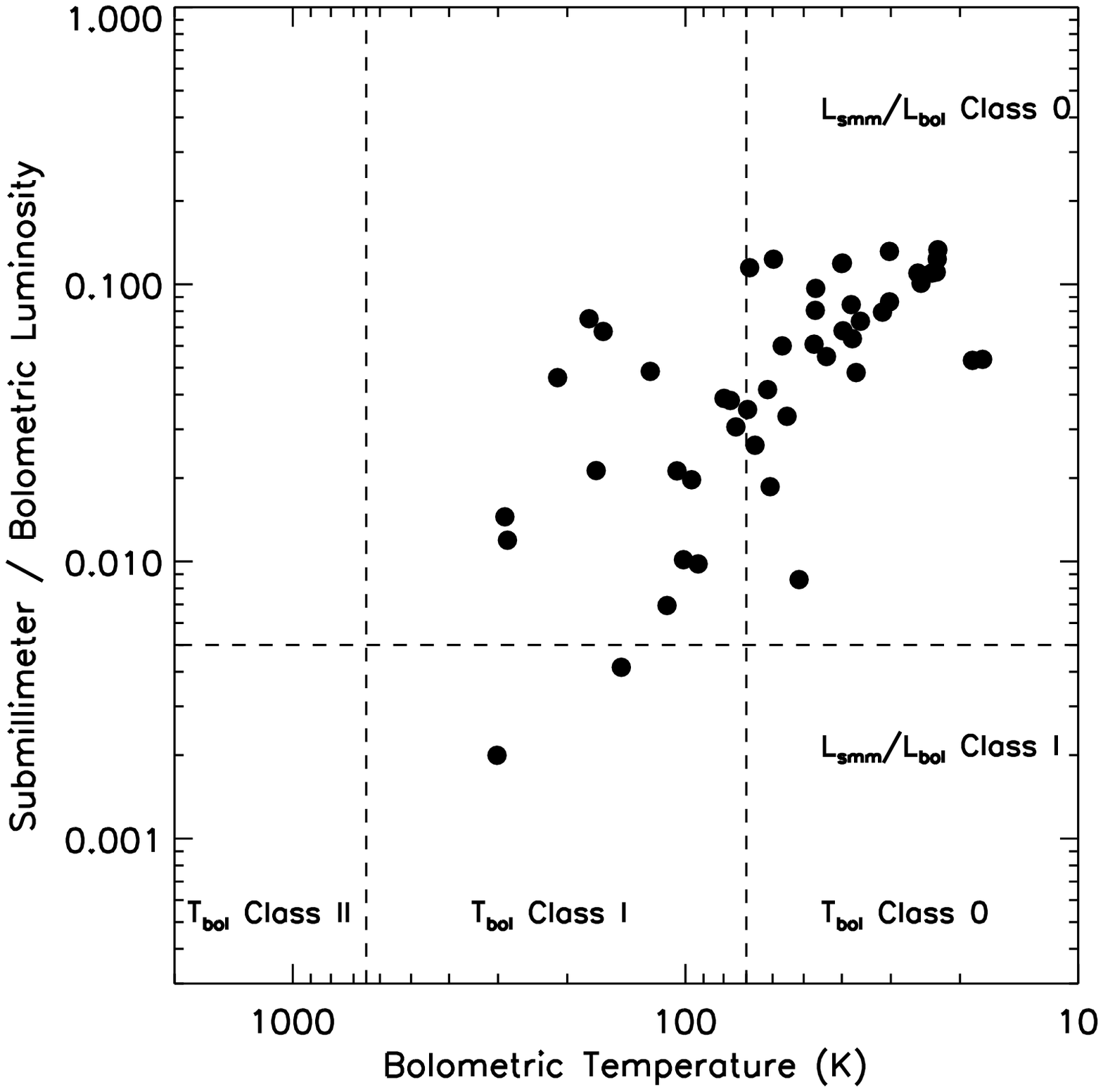}
\caption{\label{fig_classification_compare}\lsmmbol\ plotted versus \tbol, 
in both cases measured with 350 \um\ SHARC-II observations included in the 
SEDs.  The other data included in the SEDs used to calculate both values are 
the same as in Figure \ref{fig_classification_tbol}.  The vertical dashed 
lines show the Class boundaries in \tbol\ as defined by \citet{chen1995:tbol}, 
and the horizontal dashed lines show the Class boundaries in \lsmmbol\ as 
defined by \citet{andre1993:class0}.}
\end{figure}

Finally, we also investigated whether the measured flux densities of our 
science targets depended on the observing mode.  To do this, we observed 
several science sources in both observing modes. For the 23 maps observed in 
both modes, Tables \ref{tab_properties1} and \ref{tab_sources} only present 
information for the box scans; the target and source information for the 
Lissajous scans are given in Tables \ref{tab_lissajous} and 
\ref{tab_lissajous2}.  We extracted sources from these extra Lissajous maps 
and measured flux densities using the same methods as described above.  
Figures \ref{fig_lissmultiple1} -- \ref{fig_lissmultiple4} show contour maps 
overlaid on images for these extra Lissajous maps, using the maps reduced 
without the extended emission flag.  
All of the reduced FITS files used to produce these figures are available 
through the Data Behind the Figures (DBF) feature of the journal.  These maps 
are given in calibrated units of Jy beam$^{-1}$ following the calibration 
procedures described above, and both the versions with and without the 
extended flag are provided.  For the sources detected in both modes, 
Figure \ref{fig_radboxliss} plots the ratios of the flux densities calculated 
in maps observed in the box mode to those observed in the Lissajous mode 
versus the radius of each source determined from the box scan observations.  
The means and standard deviations of these ratios are $0.93 \pm 0.18$, 
$0.97 \pm 0.17$, and $1.09 \pm 0.29$ for the peak intensities, 20$''$ flux 
densities, and 40$''$ flux densities, respectively.  In all three cases the 
mean ratios are within 10\% of unity.  Given the overall calibration 
uncertainty of 25\% and the fact that the ratios show no dependence on source 
radius, we conclude that similar amounts of flux are recovered between the 
two observing modes on scales up to at least 40$''$.

\section{Sensitivity to Extended Emission}\label{sec_extended}

In addition to the detected sources listed in Table \ref{tab_sources}, 
Table \ref{tab_undetected} lists an additional 48 cores covered by our 
maps but not detected in our SHARC-II observations.  These are cores 
identified by other observations at submillimeter and millimeter wavelengths, 
based on SIMBAD\footnote{http://simbad.u-strasbg.fr} searches of the total 
area covered by our maps.  Combining the 
information from these tables, our maps cover a total of 137 protostellar 
cores, 130 of which are detected, and a total of 75 starless cores, 
34 of which are detected.  Our detection rate of 95\% for protostellar 
cores is thus much higher than our detection rate of 45\% for starless 
cores.  These results suggest that SHARC-II is well suited for identifying and 
characterizing protostellar cores, but is not ideal for studying starless 
cores. 

Of the maps observed in both the Lissajous and box scan modes, there are six 
starless cores.  Two are detected in both modes 
(L1455-IRS2E and B18-4), two are detected 
in only the box scan mode (L1544 and L694-2), and two are undetected in 
both observing modes (Perseus Bolo 107 and L1521B).  The lower detection 
rate for starless cores in the Lissajous mode (33\%) versus the box scan mode 
(45\%), coupled with the fact that two starless cores detected in the box 
scan mode are not detected in the Lissajous mode, suggest that the box 
scan mode is somewhat better suited to detecting starless cores, although 
we caution that the sample sizes are very small.

The low detection rate for starless cores is likely explained by the fact 
that, compared to protostellar cores, starless cores feature flatter 
density profiles and colder temperatures in their central regions 
\citep[e.g.,][]{wardthompson2007:ppv,evans2001:starless}.  Consequently 
starless cores exhibit 350 $\mu$m intensity profiles that are significantly 
shallower and less centrally condensed than those for 
protostellar sources, as confirmed by \citet{Wu2007} using simple, 
one-dimensional radiative transfer models.  The more extended nature of 
their emission profiles make them harder to separate from sky emission, 
thus they are less reliably detected.  To further quantify 
this effect, we examined all of the starless cores covered by our maps 
that were identified in 1.1 mm Bolocam surveys of Perseus 
\citep{1MMSurveyPers}, Ophiuchus \citep{1MMSurveyOph}, and Serpens 
\citep{1MMSurveySerp}.  Figure \ref{fig_starless} shows histograms of the 
core masses, sizes, and mean densities, as derived from the Bolocam 
observations, for both the detected and undetected populations of starless 
cores in our dataset.  The detected starless sources span nearly the full 
range of masses, FWHM angular sizes, and mean densities, indicating there is 
no one unique property that determines the detectability of the core.  

Figure \ref{fig_starless2} plots the FWHM angular size versus core mass 
for these same starless cores, again taking the properties from the 1.1 mm 
Bolocam surveys cited above.  Inspection of this figure shows that, for 
low-mass cores ($M < \sim$2 \msun), only the most compact cores are detected.  
These cores will have the steepest intensity profiles among all cores with 
such masses, allowing them to be separated from sky emission and reliably 
detected.  Only for relatively high-mass (and thus relatively bright) cores 
($M > \sim$3 \msun) are more extended cores able to be separated from sky 
emission and detected by SHARC-II observations.   Thus, starless cores 
revealed by SHARC-II surveys of star-forming regions are biased toward the 
most compact or highest mass cores, and even in cases where starless cores 
are detected, the extended nature of their intensity profiles means that 
the measured flux densities are likely lower limits to the true flux densities. 
These cautions should be kept in mind when 
interpreting the results from \citet{zhang2014}, who derive a prestellar core 
mass function based on SHARC-II observations of Ophiuchus.  This core mass 
function for detected prestellar cores may not be representative of the full 
population of such cores in this cloud, and furthermore the measured masses 
of the detected prestellar cores likely underestimate their true masses.  
Since the detectability of a starless core with SHARC-II is not a simple 
function of the total flux density of the core, but also its emission profile, 
measured upper limits for undetected cores do not necessarily represent true 
limits to the flux densities of these cores.  While we do list the 1$\sigma$ 
upper limits in Table~\ref{tab_undetected} (taken directly from 
Table~\ref{tab_properties1}), this caution should be kept in mind when 
interpreting the non-detections.

While we caution that the measured flux densities of starless cores are 
likely lower limits, the measured values for the protostellar cores are 
much more reliable.  \citet{Wu2007} used simple, one-dimensional radiative 
transfer models to show that protostellar cores exhibit 
significantly steeper intensity profiles compared to starless cores, even 
for protostars with luminosities as low as 0.1~\lsun.  These steeper intensity 
profiles lead to more compact emission that is fully recovered by SHARC-II, 
at least on scales up to the 40$''$ considered here.  This is confirmed 
by comparing the SHARC-II observations of IRAM~04191+1522 and L1521F, two 
protostars with luminosities less than 0.1~\lsun, to published radiative 
transfer models; in both cases the observed flux densities in 40$''$ diameter 
apertures agree with the models to within 2$\sigma$ 
\citep{dunham2006:iram04191,bourke2006:l1521f}.

\section{Classification of Protostars}\label{sec_classification}

Protostars are commonly classified into one of two classes, Class 0 and 
Class I, based on observational signatures that trace the underlying 
evolutionary state.  Class 0 protostars were first defined by 
\citet{andre1993:class0}, who defined such objects observationally as 
protostars emitting a relatively large fraction (greater than 0.5\%) of their 
total luminosity at wavelengths $\lambda \geq 350$ \um. Defining such 
luminosity as the submillimeter luminosity, \lsmm, 
Class 0 objects are then protostars 
with \lsmmbol~$>$~0.005.  The corresponding physical Stage 0 objects are 
young, embedded protostars with greater than 50\% of their total system mass 
still in the core \citep{andre1993:class0}.  Another quantity used to classify 
protostars is the bolometric temperature \tbol, defined by 
\citet{myers1993:tbol} as the temperature of a blackbody with the same 
flux-weighted mean frequency as the source.  By calculating \tbol\ for a large 
sample of young stars, Chen et al.~(1995) showed that Class 0 objects have 
\tbol~$<$~70~K whereas Class I protostars have \tbol~$\geq$~70~K.  Since 
\lsmm\ has historically been difficult to calculate accurately due to the 
difficulty in obtaining high-quality submillimeter data from the ground, 
particularly at 350 $\mu$m, the \tbol\ criterion introduced by Chen et al.~is 
often used instead for classifying protostars into their two Classes 
\citep[e.g.,][]{enoch2009:protostars,dunham2013:luminosities,tobin2016:vandam}. 
However, as several studies have shown that \lsmmbol\ is less sensitive to 
viewing geometry and a better tracer of underlying physical stage than \tbol\ 
\citep{andre2000:ppiv,young2005:evolmodels,dunham2010:evolmodels,frimann2015:synthetic1},
protostellar classification must be revisited as additional data becomes 
available.

With the 350 \um\ photometry presented here, we can now accurately calculate 
\citep[to within 20\%--60\%; see][for details]{dunham2008:lowlum,enoch2009:protostars} both \tbol\ and \lsmmbol\ and compare classification via the two 
quantities.  To ensure as uniform a dataset as possible, we consider only the 
protostellar cores in Perseus, and construct SEDs for each source consisting 
of {\it Spitzer Space Telescope} 3.6--70~\um\ photometry from 
\citet{evans2009:c2d} and \citet{dunham2015:gb} and Bolocam 1.1~mm photometry 
from \citet{1MMSurveyPers}.  We calculate \tbol\ twice, once without the 
SHARC-II 350~\um\ photometry included and once with it included, and we also 
calculate \lsmmbol.

Figure \ref{fig_classification_tbol} compares the two values of \tbol, 
calculated with and without the 350~\um\ photometry included.  
Leaving out the 350~\um\ photometry increases the value of \tbol, with the 
effect growing in significance with decreasing evolutionary stage (colder 
values of \tbol).  This result is explained by the fact that, for more deeply 
embedded protostars, the emission peaks at longer wavelengths, and more of 
this emission is lost when no submillimeter photometry is available.  As 
demonstrated by Figure \ref{fig_classification_tbol}, Class 0 protostars 
with very low values of \tbol\ can masquerade as more evolved objects when 
350~\um\ photometry is lacking.  These results are in qualitative agreement 
with earlier investigations by \citet{dunham2008:lowlum} and 
\citet{enoch2009:protostars}.

Figure \ref{fig_classification_compare} plots \lbolsmm\ versus \tbol\ for 
these same protostars as a way of comparing the two different classification 
methods.  Such a comparison is only possible when 350~\um\ data is available, 
since accurate calculation of \lsmm\ requires 350~\um\ data.  
With the \tbol\ Class boundaries as defined by \citet{chen1995:tbol} 
and the \lsmmbol\ Class boundaries as defined by \citet{andre1993:class0}, 
the two classification methods agree for only 65\% of the protostars considered 
(35 out of 54).  In particular, there are many protostars classified as Class I 
by \tbol\ (\tbol~$>$~70~K) but Class 0 by \lsmmbol\ (\lsmmbol~$>$~0.005).  
Revised \lsmmbol\ Class 0/I boundaries have been proposed in the literature, 
including 0.01 \citep{andre2000:ppiv,sadavoy2014:class0} and 0.03 
\citep{maury2011:aquila}.  Adopting these boundaries increases the agreement 
between classification methods to 67\% and 74\%, respectively.  

While the primary focus of this publication is to provide a catalog of 
SHARC-II 350~\um\ observations of nearby star-forming regions, these results 
demonstrate the critical role played by 350~\um\ observations in both accurate 
classification of protostars and assessing the reliability of different 
classification methods in tracing the underlying evolutionary stages of 
protostars.  A more complete investigation of protostellar classification, 
using complete far-infrared and submillimeter SEDs provided by both 
ground-based surveys such as this effort and surveys with the 
\textit{Herschel Space Observatory}, will be presented in a future paper 
(M.~M.~Dunham et al., in preparation).  

\section{Summary}\label{sec_summary}

In this paper we have presented a catalog of low-mass star-forming cores 
observed with the SHARC-II instrument at 350 $\mu$m.  Our observations have 
an effective angular resolution of 10$''$, approximately three times higher 
than observations at the same wavelength obtained with the 
{\it Herschel Space Observatory}.  A summary of our results is as follows:

 \begin{itemize}
\item We present 81 maps covering a total of 164 detected sources.  We 
tabulate basic source properties including position, peak intensity, 
flux density in fixed apertures, and radius.
\item We examine the uncertainties in the pointing model applied to all 
SHARC-II data and conservatively find that the model corrections are good 
to within $\sim$3$''$, approximately $1/3$ of the SHARC-II beam.
\item We examined the differences between the Lissajous and box scan 
observing modes.  We find that the calibration factors, beam size, and 
beam shape are similar between the two modes, and we 
also show that the same flux densities are measured when sources are observed 
in the two different modes.  Thus we conclude that there are no systematic 
effects in our catalog introduced by switching observing modes during the 
course of taking observations.
\item We find that less than half of the starless cores observed are detected 
by SHARC-II (45\% to be precise), and show that the detections are biased 
toward the most compact or highest mass starless cores.  We argue that, even 
for the detected starless cores, the measured flux densities are likely lower 
limits to the intrinsic flux densities.
\item For protostellar cores, our SHARC-II observations fully 
recover the emission, at least up to the 40$''$ scales considered here.
\item We demonstrate that the inclusion of 350 \um\ photometry significantly 
improves the accuracy of calculated values of \tbol, and enables comparison 
between two different measures of protostellar Class, \tbol\ and \lsmmbol.  
The latter can only be calculated when 350 \um\ photometry is available.
\end{itemize}

\acknowledgments
We thank the referee for helpful comments that have improved the quality 
of this publication.  
We gratefully acknowledge the assistance provided by the staff of the 
CSO in obtaining the observations presented here.  We also acknowledge 
the numerous students and postdocs from the University of Texas at Austin 
who participated in observing runs over the years, and we thank Darren Dowell 
and Attila Kov{\'a}cs for technical assistance with SHARC-II and CRUSH.  
Finally, we express our profound gratitude to everybody who played a role in 
the construction, commissioning, and operation of the CSO over its three 
decades of operation.
This work is based on data obtained with the Caltech Submillimeter Observatory 
(CSO), which was operated by the California Institute of Technology under 
cooperative agreement with the National Science Foundation (AST-0838261).  
This publication makes use of data products from the Infrared Processing and 
Analysis Center/California Institute of Technology, funded by the National 
Aeronautics and Space Administration and the National Science Foundation.  
These data were provided by the NASA/IPAC Infrared Science Archive, which is 
operated by the Jet Propulsion Laboratory, California Institute of Technology, 
under contract with NASA.  This research has made use of NASA's Astrophysics 
Data System (ADS) Abstract Service, the IDL Astronomy Library hosted by the 
NASA Goddard Space Flight Center, and the SIMBAD database operated at CDS, 
Strasbourg, France.  MMD acknowledges support from the Submillimeter Array as 
an SMA postdoctoral fellow, and from NASA ADAP grant NNX13AE54G.  NJE 
acknowledges support from NSF Grant AST-1109116 to the University of Texas at 
Austin.

\clearpage
\bibliographystyle{apj.bst}
\bibliography{Suresh_citations,source_citations}

\begin{deluxetable*}{llcccccccc}
\tabletypesize{\scriptsize}
\tablewidth{0pt}
\tablecaption{\label{tab_properties1}Observing Information}
\tablehead{
\colhead{}                    & \colhead{}                      & \colhead{Map Center}            & \colhead{Map Center} & \colhead{}         & \colhead{}                               & \colhead{}     & \colhead{1$\sigma$ Noise} & \colhead{} \\
\colhead{}                    & \colhead{Scan}                  & \colhead{R.A.}                  & \colhead{Decl.}      & \colhead{Dist.}    & \colhead{Dist.} & \colhead{Source}                            & \colhead{Obs.} & \colhead{(mJy}            & \colhead{}  \\
\colhead{\textsc{Core/Cloud}} & \colhead{Type\tablenotemark{a}} & \colhead{(2000.0)}              & \colhead{(2000.0)}   & \colhead{(pc)}     & \colhead{Ref.\tablenotemark{b}} & \colhead{Ref.\tablenotemark{c}} & \colhead{Date}  & \colhead{beam$^{-1}$)}    & \colhead{Cloud\tablenotemark{d}}                 }
\startdata
L1451-mm                          & B & 03 25 10.2 & +30 23 55.0 & 250 & 1  & 1  & 2010 Dec                & 89  & P \\          
L1448                             & B & 03 25 38.9 & +30 44 05.4 & 250 & 1  & 2  & 2008 Oct                & 151 & P \\
Perseus Bolo 18                   & L & 03 26 37.5 & +30 15 28.1 & 250 & 1  & 3  & 2007 Oct                & 104 & P \\
L1455                             & B & 03 27 41.0 & +30 12 45.0 & 250 & 1  & 2  & 2007 Oct                & 67  & P \\
IRAS 03249+2957                   & B & 03 28 00.4 & +30 08 01.3 & 250 & 1  & 4  & 2008 Oct                & 105 & P \\
Perseus Bolo 27                   & B & 03 28 33.3 & +30 19 35.0 & 250 & 1  & 3  & 2009 Dec                & 104 & P \\
Perseus Bolo 30                   & B & 03 28 39.1 & +31 06 01.8 & 250 & 1  & 3  & 2007 Oct                & 122 & P \\
NGC1333                           & B & 03 28 58.0 & +31 17 30.0 & 250 & 1  & 2  & 2008 Oct; 2009 Sep      & 170 & P \\
\enddata
\end{deluxetable*}

%\begin{landscape}
\begin{deluxetable*}{llcccccc}
\tabletypesize{\scriptsize}
\tablewidth{0pt}
\tablecaption{\label{tab_calibrators1}Calibration Scans and Factors}
\tablehead{
\colhead{Date}                                     &
\colhead{Calibrator}                                     &
\colhead{$C_{\rm peak}$}                                 &
\colhead{$C_{\rm 20}$}                                 &
\colhead{$C_{\rm 40}$} 
}
\startdata
  2003 May 17 & Mars & 1.3 & 0.021 & 0.015 \\
  2003 May 18 & Uranus & 1.2 & 0.027 & 0.022 \\
  2003 Sept 29 & Uranus & 1.7 & 0.038 & 0.031 \\
  2003 Sept 30 & Uranus & 2.3 & 0.045 & 0.034 \\
  2003 Oct 1 & Uranus & 1.8 & 0.038 & 0.031 \\
  2004 June 19 & Uranus & 2.5 & 0.049 & 0.035 \\
  2004 June 19 & Uranus & 2.3 & 0.044 & 0.032 \\
  2004 June 19 & Neptune & 1.9 & 0.044 & 0.036 \\
\enddata
\end{deluxetable*}
%\end{landscape}

%\begin{landscape}
\begin{deluxetable*}{llcccccc}
\tabletypesize{\scriptsize}
\tablewidth{0pt}
\tablecaption{\label{tab_calibrators2}Average Calibrator Factors Per Observing Run}
\tablehead{
\colhead{Date}                                     &
\colhead{$C_{peak}$}                                 &
\colhead{$C_{20}$}                                 &
\colhead{$C_{40}$} 
}
\startdata
  2003 May & 1.259 $\pm$ 0.035& 0.024 $\pm$ 0.004& 0.019 $\pm$ 0.005\\
  2003 Sept/Oct & 1.948 $\pm$ 0.352 & 0.041 $\pm$ 0.004& 0.032 $\pm$ 0.002\\
  2004 June & 2.159 $\pm$ 0.294& 0.044 $\pm$ 0.004& 0.033 $\pm$ 0.003\\
  2004 Sept & 1.597 $\pm$ 0.000& 0.034 $\pm$ 0.000& 0.019 $\pm$ 0.000\\
  2005 June & 1.409 $\pm$ 0.159& 0.034 $\pm$ 0.003& 0.024 $\pm$ 0.002\\
  2005 Nov & 1.424 $\pm$ 0.168& 0.032 $\pm$ 0.003& 0.024 $\pm$ 0.003\\
  2006 Dec & 1.243 $\pm$ 0.056& 0.029 $\pm$ 0.003& 0.023 $\pm$ 0.002\\
  2007 Oct & 1.558 $\pm$ 0.191& 0.035 $\pm$ 0.011& 0.027 $\pm$ 0.004\\
  2008 July & 1.305 $\pm$ 0.132& 0.030 $\pm$ 0.002& 0.024 $\pm$ 0.002\\
  2008 Sept/Oct & 1.333 $\pm$ 0.306& 0.023 $\pm$ 0.016& 0.018 $\pm$ 0.012\\
  2009 Sept & 1.440 $\pm$ 0.220& 0.033 $\pm$ 0.005& 0.027 $\pm$ 0.004\\
  2009 Dec & 1.177 $\pm$ 0.135& 0.028 $\pm$ 0.003& 0.022 $\pm$ 0.002\\
  2010 July/Aug & 1.362 $\pm$ 0.213& 0.031 $\pm$ 0.004& 0.025 $\pm$ 0.004\\
  2010 Dec & 1.377 $\pm$ 0.053& 0.030 $\pm$ 0.002& 0.022 $\pm$ 0.001\\
  All Runs & 1.462 $\pm$ 0.258& 0.033 $\pm$ 0.007& 0.025 $\pm$ 0.004
  %2008 Sept AND & & & \\
  %2009 Sept/Dec & 1.405 $\pm$ 0.215& 0.032 $\pm$ 0.005 & 0.025 $\pm$ 0.004 
\enddata
\end{deluxetable*}
%\end{landscape}

%\begin{landscape}
\begin{deluxetable*}{llccccccc}
\tabletypesize{\scriptsize}
\tablewidth{0pt}
\tablecaption{\label{tab_sources}Source Properties}
\tablehead{
\colhead{}                                     &
\colhead{}                                     &
\colhead{Peak}                                 &
\colhead{Peak}                                 &
\colhead{}                           &
\colhead{}                           &
\colhead{}                                     &
\colhead{}                               &
\colhead{}                                    \\
\colhead{}                                     &
\colhead{}                                     &
\colhead{R.A.}                                 &
\colhead{Decl.}                                &
\colhead{Radius}                &
\colhead{S$_{\rm peak}$}               &
\colhead{S$_{\rm 20"}$}       &
\colhead{S$_{\rm 40"}$}                       &              
\colhead{}                                          \\
\colhead{\textsc{Source}}                      &
\colhead{\textsc{Core/Cloud}\tablenotemark{a}} &
\colhead{(2000.0)}                             &
\colhead{(2000.0)}                             &
\colhead{(arcsec)}                             &
\colhead{(Jy/Beam)}                             &
\colhead{(Jy)}                                   &
\colhead{(Jy)}                                   &
\colhead{Status\tablenotemark{a}}                             
}    

\startdata    
   L1451-mm SMM1 & L1451-mm &  03 25 08.3 &  +30 24 28.7   &  13.8 &  0.5 $\pm$ 0.1   &    1.3 $\pm$ 0.3 &    3.4 $\pm$ 1.1 & S\\
   L1451-mm & L1451-mm &  03 25 10.4 &  +30 23 55.7    & 13.4 &  0.8 $\pm$ 0.2    &   2.1 $\pm$ 0.5 &   4.1 $\pm$ 1.3 & P\\
L1448-IRS2 & L1448 & 03 25 22.7 &  +30 45 12.1  &  15.2 &   10.3 $\pm$ 2.2      & \nodata   &      \nodata & P \\
L1448-IRS2E & L1448 &  03 25 25.8  & +30 44 57.1  &   20.5 &  2.2 $\pm$ 0.5   &  4.7 $\pm$ 1.2  &  12.3 $\pm$ 4.0 & P \\
Perseus Bolo 8-A & L1448 & 03 25 35.9  & +30 45 33.2   &  23.0 & 13.6 $\pm$ 2.9     & 23.3 $\pm$ 6.0 &   59.2 $\pm$ 19.3 & S \\
L1448-N(A) & L1448 & 03 25 36.6 &  +30 45 15.1   &  17.6 & 27.1 $\pm$ 5.7 &   42.1 $\pm$ 10.8  &  71.2 $\pm$ 23.4 & P\\
L1448-MM & L1448 & 03 25 39.1   & $+$30 44 03.2  &   20.5 & 22.9 $\pm$ 4.8  & 30.3 $\pm$ 7.8   &    45.5 $\pm$ 14.9 & P \\
Perseus Bolo 13 & L1448 & 03 25 49.5 &  +30 45 03.1  &   16.2 &  0.6 $\pm$ 0.1   & 1.6 $\pm$ 0.4  &     5.1 $\pm$ 1.7 & S \\
\enddata
\tablenotetext{a}{A ``P'' indicates that this core is a protostellar core. An ``S'' indicates that this core is a starless core.}
\end{deluxetable*}
%\end{landscape}

%\begin{landscape}
\begin{deluxetable*}{llcccccc}
\tabletypesize{\scriptsize}
\tablewidth{0pt}
\tablecaption{\label{tab_calibrators3}Box Calibration Scans and Factors}
\tablehead{
\colhead{Date}                                     &
\colhead{Calibrator}                                     &
\colhead{$C_{\rm peak}$}                                 &
\colhead{$C_{\rm 20}$}                                 &
\colhead{$C_{\rm 40}$} 
}
\startdata
  2010 July 23 & Uranus & 1.5 & 0.030 & 0.024 \\
  2010 July 23 & Uranus & 1.3 & 0.028 & 0.021 \\
  2010 July 23 & Uranus & 1.8 & 0.036 & 0.025 \\
  2010 July 23 & Uranus & 1.7 & 0.035 & 0.025 \\
  2010 July 23 & Uranus & 1.7 & 0.035 & 0.025 \\
  2010 July 23 & Uranus & 1.3 & 0.027 & 0.022 \\
  2010 July 23 & Uranus & 1.4 & 0.029 & 0.023 \\
  2010 July 23 & Uranus & 1.4 & 0.030 & 0.024 \\
  2010 July 23 & Uranus & 1.5 & 0.031 & 0.024 \\
  2010 July 23 & Uranus & 1.5 & 0.031 & 0.024 \\
  2010 July 23 & Uranus & 1.6 & 0.033 & 0.026 \\
  2010 July 23 & Uranus & 1.6 & 0.033 & 0.025 \\
  2010 July 31 & Neptune & 1.7 & 0.037 & 0.029 \\
  2010 July 31 & Neptune & 1.7 & 0.037 & 0.029 \\
  2010 July 31 & Neptune & 1.7 & 0.038 & 0.030 \\
  2010 July 31 & Neptune & 1.7 & 0.038 & 0.030 \\
  2010 July 31 & Neptune & 1.7 & 0.038 & 0.030 \\
  2010 Dec 6 & Uranus & 1.8 & 0.043 & 0.031 \\
  2010 Dec 6 & Uranus & 1.9 & 0.037 & 0.026 \\
  2010 Dec 6 & Uranus & 1.8 & 0.036 & 0.025 \\
  2010 Dec 6 & Uranus & 1.6 & 0.033 & 0.023 \\
  2010 Dec 6 & Uranus & 1.8 & 0.035 & 0.024 \\
  2010 Dec 6 & Uranus & 1.8 & 0.035 & 0.024 \\
  2010 Dec 7 & Uranus & 1.5 & 0.030 & 0.021 \\
  2010 Dec 7 & Uranus & 1.6 & 0.032 & 0.022 \\
  2010 Dec 7 & Uranus & 1.7 & 0.033 & 0.023 \\
  2010 Dec 7 & Uranus & 1.7 & 0.034 & 0.022 \\
  2010 Dec 7 & Uranus & 1.7 & 0.035 & 0.024 
\enddata
\end{deluxetable*}
%\end{landscape}

\begin{deluxetable*}{llccc}
\tabletypesize{\scriptsize}
\tablewidth{0pt}
\tablecaption{\label{tab_lissajous}Lissajous Observing Information for Sources Observed in Both Scan Modes}
\tablehead{
\colhead{}                                     &
\colhead{Map Center}                                 &
\colhead{Map Center}                                 &
\colhead{}                                     &
\colhead{1$\sigma$ Noise}                                    \\
\colhead{}                                     &
\colhead{R.A.}                                 &
\colhead{Decl.}                                &
\colhead{Obs.}       &
\colhead{(mJy)}                                     \\
\colhead{\textsc{Core/Cloud}}                      &
\colhead{(2000.0)}                             &
\colhead{(2000.0)}                             &
\colhead{Date}                                &
\colhead{beam$^{-1}$}                    
}          
L1455 & 03 27 41.0 & +30 12 45.0 & 2005 Nov & 137 \\
Perseus Bolo 25 & 03 28 32.5 & +31 11 04.9 & 2006 Dec & 40 \\
NGC 1333 IRAS 4& 03 29 13.5 & +31 13 58.1 & 2007 Oct& 341 \\
Perseus Bolo 52 & 03 29 17.2 & +31 27 46.4 & 2007 Oct & 98 \\
IRAS 03282+3035 & 03 31 20.4 & +30 45 24.7 & 2005 Nov & 169 \\
Perseus Bolo 79 & 03 33 14.4 & +31 07 10.9 & 2006 Dec & 358 \\
Perseus Bolo 102 & 03 43 51.0 & +32 03 07.9 & 2007 Oct & 195 \\
HH 211-MM & 03 43 56.8 & +32 00 50.2 & 2005 Nov & 226 \\
\enddata
\end{deluxetable*}

%\begin{landscape}
\begin{deluxetable*}{llccccccc}
\tabletypesize{\scriptsize}
\tablewidth{0pt}
\tablecaption{\label{tab_lissajous2}Lissajous Source Properties for Sources Observed in Both Scan Modes}
\tablehead{
\colhead{}                                     &
\colhead{}                                     &
\colhead{Peak}                                 &
\colhead{Peak}                                 &
\colhead{}                           &
\colhead{}                           &
\colhead{}                                     &
\colhead{}                               &
\colhead{}                                    \\
\colhead{}                                     &
\colhead{}                                     &
\colhead{R.A.}                                 &
\colhead{Decl.}                                &
\colhead{Radius}                &
\colhead{S$_{\rm peak}$}               &
\colhead{S$_{\rm 20"}$}       &
\colhead{S$_{\rm 40"}$}                       &              
\colhead{}                                          \\
\colhead{\textsc{Source}}                      &
\colhead{\textsc{Core/Cloud}\tablenotemark{a}} &
\colhead{(2000.0)}                             &
\colhead{(2000.0)}                             &
\colhead{(arcsec)}                             &
\colhead{(Jy/Beam)}                             &
\colhead{(Jy)}                                   &
\colhead{(Jy)}                                   &
\colhead{Status\tablenotemark{a}}                             
}    

\startdata    
  L1455-IRS5 & L1455 &  03 27 38.5 &  +30 13 59.2   &  17.2 &  2.7 $\pm$ 0.7  &   7.5 $\pm$ 2.7    &  18.6 $\pm$ 6.6 & P\\
 L1455-IRS1 & L1455 &  03 27 39.3 &  +30 13 05.7   &  13.8 &  5.2 $\pm$ 1.3   &  12.2 $\pm$ 4.3   &     20.4 $\pm$ 7.2 & P\\
  L1455-IRS4 & L1455 &  03 27 43.1  & +30 12 32.2   &  14.1 &  2.9 $\pm$ 0.7   &    5.9 $\pm$ 2.1  &  8.9 $\pm$ 3.1 & P\\
L1455-IRS2E & L1455 &  03 27 48.3 &  +30 12 09.7    & 16.7 &  1.2 $\pm$ 0.3   &    3.1 $\pm$ 1.1    &   5.7 $\pm$ 2.0 & S\\
   Perseus Bolo 25 & Perseus Bolo 25 &  03 28 32.7  &  +31 11 08.6   &   14.1  & 0.7 $\pm$ 0.2   &   1.5 $\pm$ 0.5   &   2.4 $\pm$ 0.8 & P\\
   NGC 1333 IRAS 4A & NGC 1333 &  03 29 10.6  &  +31 13 36.3   &  20.0 & 31.4 $\pm$ 7.9  &   54.9  $\pm$ 19.4 &   \nodata &P\\
   NGC 133 IRAS 4B & NGC 1333 &  03 29 12.0 &   +31 13 12.3   &  11.0 & 28.0 $\pm$ 7.0   &   38.4  $\pm$ 13.6  &   \nodata & P\\
   NGC 1333 IRAS 4C & NGC 1333 &  03 29 13.6  &  +31 14 01.8  &  17.5  & 4.6 $\pm$ 1.2  &   8.8  $\pm$ 3.1 &   17.1 $\pm$ 6.1 & P\\
\enddata
\tablenotetext{a}{A "P" indicates that this core is a protostellar core. An "S" indicated that this core is a starless core.}
\end{deluxetable*}
%\end{landscape}

\clearpage
%\begin{landscape}
\begin{deluxetable*}{llcccc}
\tabletypesize{\scriptsize}
\tablewidth{0pt}
\tablecaption{\label{tab_undetected}Undetected Sources}
\tablehead{
\colhead{}                                     &
\colhead{}                                     &
\colhead{R.A.}                                 &
\colhead{Dec.}                                 &
\colhead{}                                     &
\colhead{1$\sigma$ Noise}                      \\
\colhead{\textsc{Source}}                      &
\colhead{\textsc{Core/Cloud}\tablenotemark{a}} &
\colhead{(2000.0)}                             &
\colhead{(2000.0)}                             &
\colhead{Status\tablenotemark{a}}              &                        
\colhead{(\mjybeam)}                          
}    

\startdata    
Perseus Bolo 11   & L1448             & 03 25 46.0 & $+$30 44 10.3 & S & 151 \\
IRAS 03249+2957   & IRAS 03249+2957   & 03 28 00.4 & $+$30 08 01.3 & P & 105 \\
Perseus Bolo 26   & NGC1333           & 03 28 32.4 & $+$31 04 42.9 & S & 170 \\
Perseus Bolo 33   & Perseus Bolo 30   & 03 28 42.6 & $+$31 06 12.2 & S & 122 \\
Perseus Bolo 34   & NGC1333           & 03 28 45.9 & $+$31 15 19.8 & S & 170 \\
Perseus Bolo 35   & NGC1333           & 03 28 48.4 & $+$31 16 01.9 & S & 170 \\
Perseus Bolo 37   & NGC1333           & 03 28 52.1 & $+$31 18 07.9 & S & 170 \\
Perseus Bolo 51   & NGC1333           & 03 29 17.0 & $+$31 12 25.9 & S & 170 \\
\enddata
\tablenotetext{a}{A ``P'' indicates that this core is a protostellar core. An ``S'' indicates that this core is a starless core.}
\end{deluxetable*}
%\end{landscape}

\end{document}